\documentclass[twocolumn]{aastex62}

\usepackage{graphicx}
\usepackage{natbib}
\usepackage{multirow}
\usepackage{textcomp}
\usepackage{longtable}
\usepackage{ulem}
\usepackage{chngcntr}
\usepackage{hyperref}
\usepackage{changepage}
\usepackage{amsmath}
\usepackage{rotating}

\newcommand{\HI}{\mbox{H{\sc i}}}
\newcommand{\kms}{km s$^{-1}$}
\newcommand{\degree}{$^{\circ}$}

\newcommand{\flux}{F$_{\HI}$}
\newcommand{\vhel}{v$_{hel}$}
\newcommand{\wfifty}{w$_{50}$}
\newcommand{\wtwenty}{w$_{20}$}
\newcommand{\distance}{D$_{LG}$}
\newcommand{\mass}{M$_{\HI}$}

\begin{document}

\title{The Arecibo L-band Feed Array Zone of Avoidance (ALFAZOA) Shallow Survey}

\author{M. Sanchez-Barrantes}
\affil{Department of Physics and Astronomy, University of New Mexico, 1919 Lomas Blvd. NE, Albuquerque, NM 87131, USA}
\affil{National Radio Astronomy Observatory, P.O. Box 0, Socorro, NM 87801, USA}

\author{P. A. Henning}
\affil{Department of Physics and Astronomy, University of New Mexico, 1919 Lomas Blvd. NE, Albuquerque, NM 87131, USA}

\author{T. McIntyre}
\affil{The Pew Charitable Trusts, 901 E Street NW, Washington, DC 20004 \footnote{Research completed at UNM before starting at Pew}}

\author{E. Momjian}
\affil{National Radio Astronomy Observatory, P.O. Box 0, Socorro, NM 87801, USA}

\author{R. Minchin}
\affil{Infrared Astronomy/USRA, NASA Ames Research Center, MS 232-12, Moffett Field, CA 94035, USA}

\author{J.L Rosenberg}
\affil{George Mason University, Department of Physics and Astronomy, 4400 University Drive, Fairfax, VA 22030}

\author{S. Schneider}
\affil{University of Massachusetts, Department of Astronomy, 619E LGRT-B, 01003, Amherst, MA, USA}

\author{L. Staveley-Smith}
\affil{International Centre for Radio Astronomy Research, University of Western Australia, Crawley, WA 6009, Australia}
\affil{ARC Centre of Excellence for All Sky Astrophysics in 3 Dimensions (ASTRO 3D)}

\author{W. van Driel}
\affil{GEPI, Observatoire de Paris, PSL Research University, CNRS, 5 place Jules Janssen, 92190 Meudon, France}

\author{M. Ramatsoku}
\affil{INAF-Osservatorio Astronomico di Cagliari, Via della Scienza 5, 09047 Selargius (CA), Italy}

\author{Z. Butcher}
\affil{University of Massachusetts, Department of Astronomy, 619E LGRT-B, 01003, Amherst, MA, USA}

\author{E. Vaez}
\affil{International Centre for Radio Astronomy Research, University of Western Australia, Crawley, WA 6009, Australia}

\begin{abstract}

The Arecibo L-band Feed Array Zone of Avoidance (ALFAZOA) Shallow Survey is a blind \HI~survey of the extragalactic sky behind the northern Milky Way conducted with the ALFA receiver on the 305m Arecibo Radio Telescope. ALFAZOA Shallow covered 900 square degrees at full sensitivity from 30\degree $\leq~l~\leq$ 75\degree~and $ \lvert b \lvert $ $\leq$ 10\degree~and an additional 460 square degrees at limited sensitivity at latitudes up to 20\degree. It has an rms sensitivity of 5 -7 mJy and a velocity resolution of 9 - 20.6 \kms, and detected 403 galaxies out to a recessional velocity of 12,000 \kms, with an angular resolution of 3.4' and a positional accuracy between 0.2' and 1.7'. The survey is complete above an integrated line flux \flux~=~2.0 Jy \kms~for half the survey, and above \flux~=~2.8 Jy \kms~for the other half. 

Forty-three percent of the ALFAZOA \HI~detections have at least one possible optical/NIR counterpart in the literature, and an additional 16\% have counterparts that only included previous \HI~measurements. There are fewer counterparts in regions of high extinction and for galaxies with lower \HI~mass. Comparing the results of the survey to the predictions of \cite{Erdogdu06}, and using their nomenclature, ALFAZOA confirms the position and extent in the ZOA of the C7, C$\zeta$, Pegasus, Corona Borealis and Delphinus structures, but not of the Cygnus void. Two new structures are identified, both connected to the C7 overdensity; one extends to ~35\degree, and the other crosses the ZOA.

\end{abstract}


\section{Introduction} \label{sec:intro}

Observations of extragalactic objects behind the Milky Way are limited in the optical and near-infrared (NIR) by high extinction levels due to dust and confusion due to stars from our Galaxy. This region is known as the Zone of Avoidance (ZOA) due to the historically low detection rate of galaxies in the region. The ZOA blocks about 20\% of the extragalactic sky at optical wavelengths, and about 10\% at near-infrared wavelengths. The most homogeneous NIR wide-angle redshift survey, the Two Micron All Sky Survey (2MASS) Redshift Survey, 2MRS (\citealt{Huchra12}), has a gap of $\pm$ 5\degree~to 8\degree~in Galactic latitude around the Galactic plane. In contrast to this, observations in the 21cm line of neutral hydrogen (\HI) are not affected by extinction and galaxies at recessional velocities separate from Galactic \HI~ can be easily identified.

\begin{table*}[htbp]
\caption{Overview of ALFAZOA Shallow Parameters}
\begin{center}
\begin{tabular}{|ccc|}
\hline \hline
Parameter&Field&Value\\
\hline
Galactic latitude&&$ \lvert b \lvert $ $\leq$ 10\degree\\
Galactic longitude&& 30\degree $\leq l \leq$ 75\degree\\
Velocity range&B, C&-1000 $\leq$ \vhel $\leq$ 11,500 \kms\\
&A, D&-1000 $\leq$ \vhel $\leq$ 10,500 \kms\\
Velocity resolution&B, C&9 \kms\\
&A, D&20.6 \kms\\
Survey size&complete& 1200 square degrees\\
&limited sensitivity& 160 square degrees\\
Integration time &&8s per beam\\
Beam size at FWHM&&3.4 arcmin\\
RMS noise per channel&B, C (9 \kms~channels)&5.4 mJy\\
&A, D (20.6 \kms~channels) &7 mJy\\
\hline
\end{tabular}
\end{center}
\label{summary}
\end{table*}%

The missing information from ``all-sky" redshift and peculiar velocity surveys in the ZOA (e.g. 2MRS; \citealt{Huchra12} and 2MASS Tully Fisher; \citealt{Masters08}) is an obstacle to our understanding of the dynamics of the Local Group (LG) and of the large-scale structure (LSS) of the universe. This problem is usually worked around by using statistical interpolation of the mass distribution next to the ZOA, such as the reconstruction of the local Universe done by \cite{Erdogdu06}. However, we have to consider the possibility that there are unpredicted structures in this region which can have an influence on the dynamics of the LG, such as the predictions from \cite{Loeb08}. \cite{Loeb08} studied the motion of the LG by comparing the dipoles of the cosmic microwave background and that expected from the gravitational acceleration imparted on the LG as measured from the 2MRS, and found a discrepancy they believed could be explained by undiscovered mass in the ZOA. Moreover, new additions to the LSS in the ZOA are still being made, such as the recently discovered Vela Supercluster (\citealt{kraan15}, \citealt{Kraan17}). Missing galaxies impact the interpretation of the dynamics of the Local Universe, and a complete velocity survey that includes the ZOA is necessary to understand this, as well as the observed CMB dipole discrepancy, or the cosmic flow fields. 

The survey presented in this paper is the Arecibo L-Band Feed Array Zone of Avoidance (ALFAZOA) Survey, a blind, wide-angle \HI~survey with observations carried out by the 305m Arecibo Telescope in the ZOA. ALFAZOA is made up of two phases: Shallow and Deep. The Shallow Survey, which is the focus of this paper, has an rms noise of 5 -7 mJy (\citealt{Henning10}) at a velocity resolution of 9.0-20.6 \kms. It covers the region of the sky between galactic longitudes by 30\degree $\leq~l~\leq$ 75\degree and galactic latitudes $ \lvert b \lvert $ $\leq$ 10\degree, from $-$1000 \kms~to 12,000 \kms~in heliocentric velocity (Figure \ref{coverage}). Although it extends in parts beyond $b$ = $\pm$10\degree, it is only within this range that the survey is complete and has optimal sensitivity. The Deep Survey is ongoing and five times more sensitive than the Shallow, and covers a very narrow strip closer to the Galactic Center, as well as a strip in the outer galaxy part of the northern ZOA.
 
There are other ongoing and completed \HI~surveys with a focus on the ZOA. Most notably, in the southern hemisphere there is the blind Parkes \HI~Zone of Avoidance Survey (HIZOA; \citealt{Henning00}, \citealt{Donley05}, \citealt{Stavely16}), with an rms noise of 6 mJy and a velocity resolution of 27 \kms, that covered the sky from 196\degree~$<~l~<$ 52\degree~and $ \lvert b \lvert $ $<$ 5\degree. In the north, the only other blind survey is the all-northern hemisphere Effelsberg Bonn \HI~Survey (EBHIS; \citealt{Kerp11}), which is underway and will have an rms noise of about 13 mJy and an angular resolution of 9', at a channel width of 18 \kms. Follow-up observations of a hundred EBHIS ZOA sources from the first catalog are underway at Nan\c{c}ay (Shr{\"o}der et al., in prep.). In comparison, ALFAZOA Shallow has a sensitivity of 5-7mJy and a better angular resolution (3.4'). In addition to these blind surveys, there have been complementary surveys of small areas in the ZOA (e.g.\citealt{Ramatsoku16}; a deep interferometric survey with better angular resolution), and targeted observations of a few hundred 2MASS-selected galaxies which reveal structure behind the northern ZOA (Nan{\c c}ay: eg. \citealt{vanDriel09}, \citealt{Kraan18}) and southern ZOA (Parkes: \citealt{Said16} ).

In Section \ref{sec:obs} we describe the observations for the ALFAZOA Shallow survey. Section \ref{sec:data} describes the method used to produce the spectral cubes, section \ref{sec:search} how the cubes were searched for galaxies and the resulting catalog is presented in \ref{sec:over}. Section \ref{sec:counter} looks into potential counterparts from other surveys and wavelengths, and in Section \ref{sec:selfun} details of the survey performance are examined. Section \ref{sec:LSS}~discusses the large-scale distribution of the galaxies detected in this survey and their relation with existing LSS. Finally, the conclusions are presented in section \ref{sec:concl}.


\section{Observations} \label{sec:obs}

\begin{figure} [t]
\centering
\includegraphics[width=0.45\textwidth]{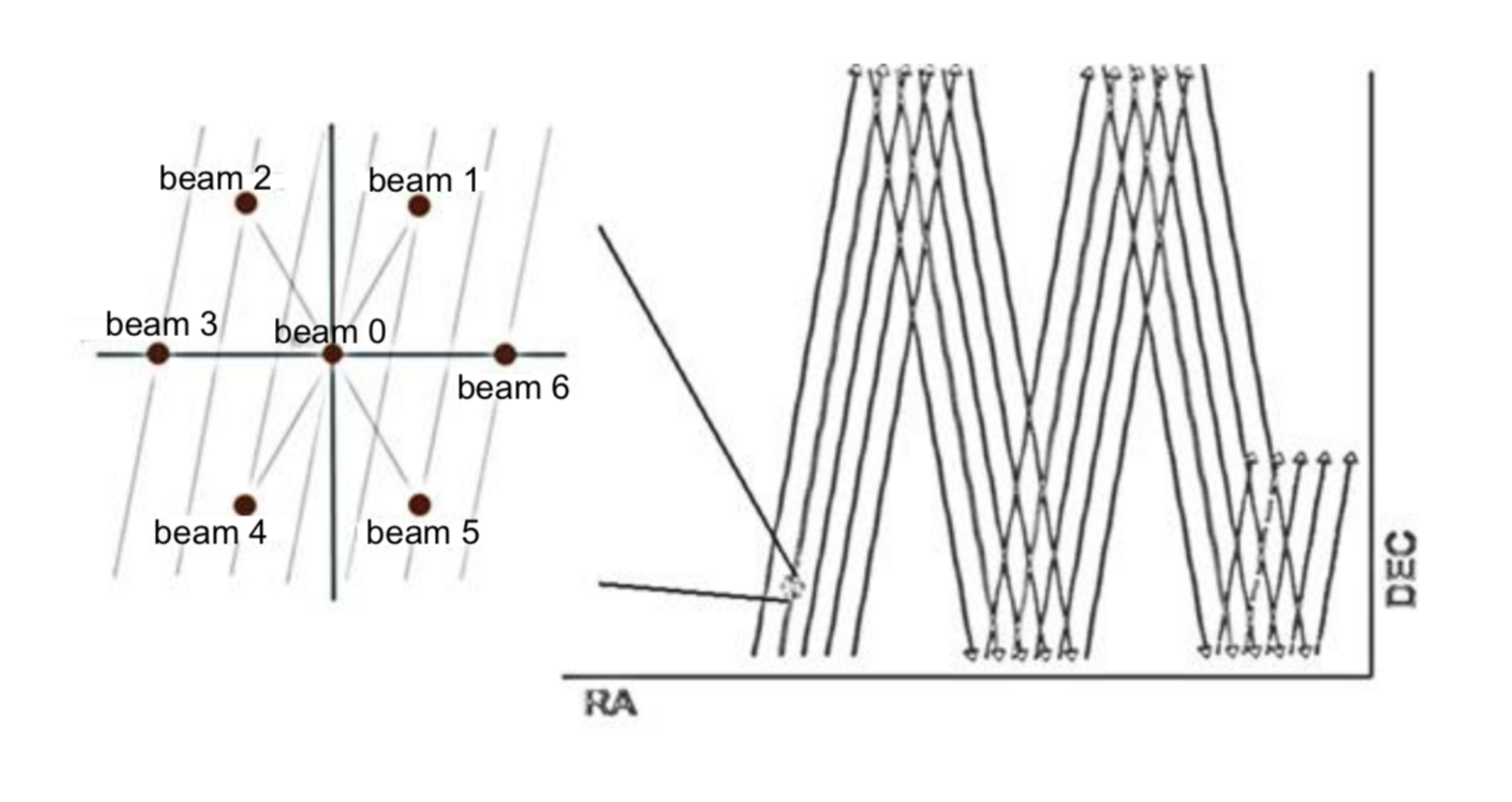}
\caption{Illustration of the meridian nodding mode observing technique. Each observing scan makes a W-shape on the sky, with seven equally spaced beam scans separated by 1.83'. On the right we can see the tracks made in the sky each night by the telescope, which were separated so that the distance between beam 6 and beam 3 of subsequent days is the same 1.83' between each beam. (Image is 3.2 of \citealt{McIntyreTh})}\label{lambda}
\end{figure}

The observations described here were carried out with the Arecibo L-band Feed Array (ALFA) on the 305m telescope located in Arecibo, Puerto Rico, from May 2008 to August 2009. The total observing time was 322 hours, and the survey was observed simultaneously with other astronomical projects. The ALFA receiver consists of seven independent beams with two orthogonal linear polarizations each, arranged with one central beam surrounded by six outer beams in a hexagonal pattern (see Figure \ref{lambda}). At 1.4 GHz, the FWHM of each beam is approximately 3.4' , and the mean system temperature is 30K. More details about ALFA can be found in \cite{Giovanelli05}.

\begin{figure} 
\includegraphics[width=0.45\textwidth]{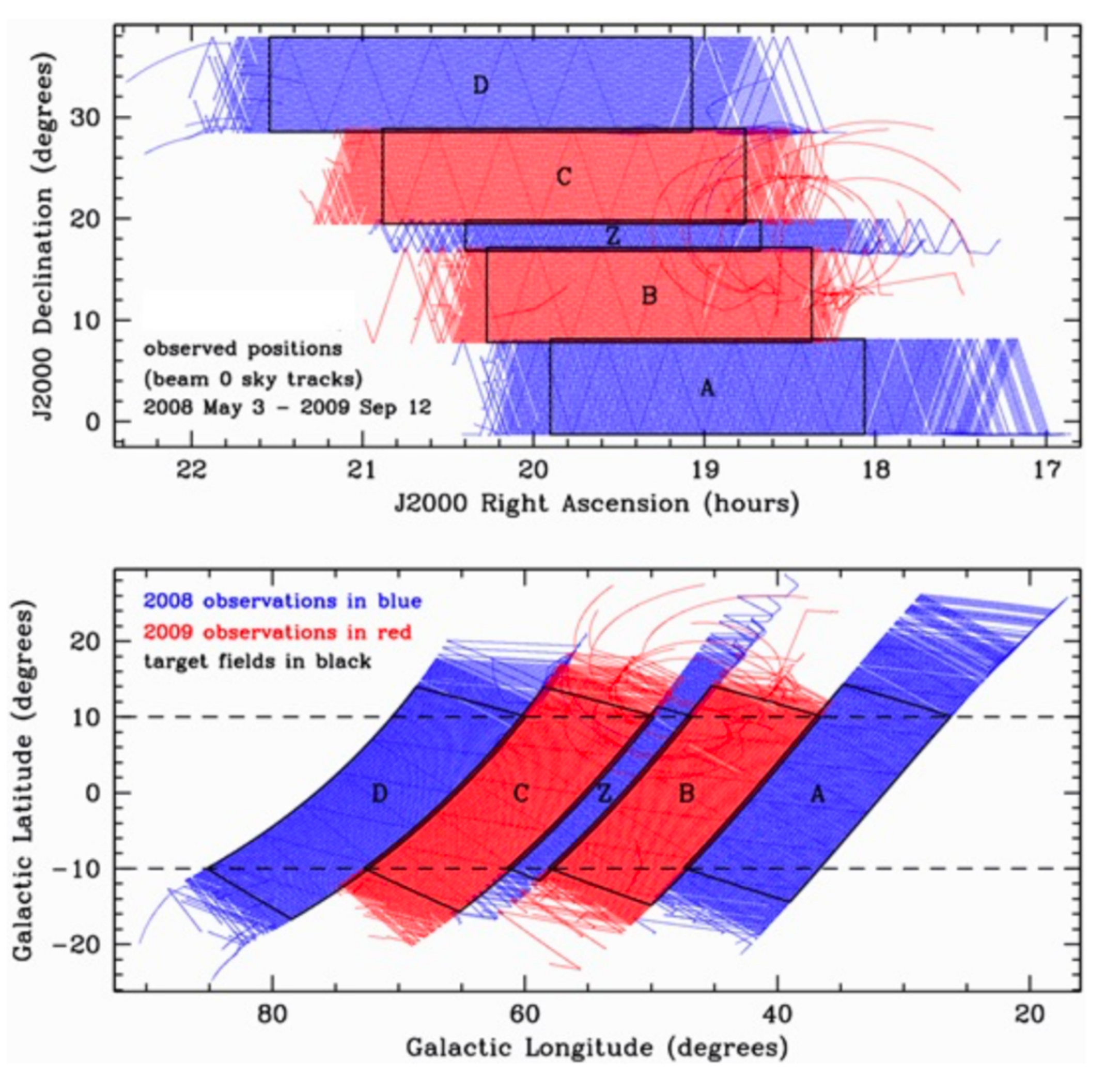}
\caption{The scan pattern of all observations made for the ALFAZOA Survey in Equatorial coordinates (top) and Galactic coordinates (bottom). The target fields are outlined in black. Fields A and D were observed in 2008 with the WAPP spectrometer and fields B and C were observed in 2009 with the Mock spectrometer. See Section \ref{sec:data} for details. Image courtesy of commensal partner I-GALFA. The I-GALFA region Z is not included in this paper.}
\label{coverage}
\end{figure}

The ALFAZOA Shallow Survey covers about 1200 square degrees at optimal sensitivity (full and complete coverage) and an additional 160 square degrees at limited sensitivity between 30\degree~$\leq l \leq$ 75\degree~and $ \lvert b \lvert $ $\leq$ 15\degree (Figure \ref{coverage}). The survey presented here covers four different areas of the sky: A, B, C and D, each covering a different region of the sky (Figure \ref{coverage}). They were observed in meridian nodding mode, as described in detail in \cite{Henning10}. The motion of the receiver (Figure \ref{lambda}) and the overlapping of each day's observations resulted in an effective integration time for this survey of 8s per beam.

The Shallow survey used two different spectrometer backends at different times over the course of the observations; the Wideband Arecibo Pulsar Processor (WAPP) and the Mock spectrometer. Fields A and D were observed in 2008 using the WAPP spectrometer with 2048 channels and covering a frequency range from 1330 MHz to 1430 MHz. The integrations were done every 200 ms and a low noise diode was fired for 100 ms every integration because of the calibration requirements of commensal partners. The resulting channel spacing was 10.3 \kms, and the searchable range for the 21cm \HI~line in heliocentric velocity (\vhel) was $-$1000 \kms and 10,500 \kms. 

Fields B and C were observed in 2009 using the Mock spectrometer with 8192 channels and a bandwidth from 1225 MHz to 1525 MHz. The Mock spectrometer recorded 1s integrations, and a low noise diode was fired during the observations to meet the calibration requirements of commensal partners, but was not injected in the Mock data set.  The resulting channel spacing was 4.5 \kms and the searchable velocity range for these two fields was from $-$1000 \kms~to 11,500 \kms~because only the high-frequency sub-band was used. Both WAPP and Mock spectrometers sampled a high flux noise diode for 3s integrations at the top and bottom of the nodding scans. 

A summary of the survey parameters is listed in Table \ref{summary}, where the velocity resolution is after Hanning smoothing as described in Section \ref{sec:data}.

\bigskip
\bigskip
\bigskip

\section{Data Reduction} \label{sec:data} 

\subsection{B and C fields}

The data for B and C fields was processed in the same way described in \cite{Henning10}. Using LiveData (\citealt{Barnes01}), a median bandpass was calculated from each beam and polarization and was removed, the system temperature calibration was applied, and the data was Doppler corrected. Hanning smoothing was applied to the final spectrum, resulting in a velocity resolution of 9 \kms.

The calibrated spectral data was then gridded into overlapping spectral data cubes using Gridzilla (\citealt{Barnes01}), with a pixel size of 1 arcmin, where the value of each pixel is the median of the spectra that lie within 1.5 arcmin of the center of that pixel, which helps to mitigate transient RFI (\citealt{Henning10}).More details on the data reduction for B and C fields can be found at \cite{McIntyreTh}.

\subsection{A and D fields} \label{sec:ad}

The data for A and D fields had to be processed differently than the data from B and C. Because of the science goals of the other astronomical project that was being run together with ALFAZOA, the data taken with the WAPP spectrometer was done in pulsar mode, which resulted in binary files with a limited set of headers, and data that were not compatible with LiveData. The data reduction process used on A and D field was developed in IDL by the ALFAZOA team. Details can be read in \cite{McIntyreTh}.

Unlike in B and C, half of every 200 ms integration had the low flux noise diode on due to calibration requirements for the commensal project, and this had to be extracted from the bandpass by subtracting the half of the spectra where the diode is on from the half where it is off, and dividing it by the value when it is off. A median of all occurrences within a scan was used to create a low flux noise diode bandpass that is subtracted from every instance where it was fired. 

The spectral data were then bandpass corrected in IDL in the same way as was described for B and C. The calibrated spectral data was then made into spectral cubes of the same size as those from B and C fields.

The complicated data reduction process of A and D fields resulted in problems with the flux calibration of these fields, and this required recalibration of the flux density scale. We corrected the flux density scale by comparing the integrated \HI~line flux from our observations to measurements in the literature for 12 galaxies in our field that were previously observed by the Arecibo telescope. We also compared three galaxies which are in the region of overlap between fields A/B, and C/D in the same manner. The data from the previously observed galaxies were smoothed to the same velocity resolution as the A and D field data, and the integrated flux of each was measured with the Spectral Analysis Tool (SPLAT) \citep{Skoda14}. 

The flux density scale factor was calculated for each galaxy as the ratio of integrated flux from the literature divided by our integrated ALFAZOA flux. There were fourteen galaxies inside the region of A and D with Arecibo \HI~measurements, but two of those were removed from the sample due to the bad data quality of their previous observations. A weighted mean scale factor was derived from the final twelve galaxies, taking into account the uncertainty for each individual galaxy. This value was applied to all of the A and D cubes using the program MATHS in the MIRIAD radio interferometry data reduction package (\citealt{Sault95}), and the resulting values are those presented in this paper. The scale factor was applied uniformly to all data, so that both the signal and the noise were modified with this change. The standard deviation of the scale factor introduces an uncertainty in the measurements of this field of 4\% to the given measurement.

The A and D fields have higher noise in the data compared to the noise measured in B and C fields (7.0 mJy and 5.4 mJy, respectively). The edges of the sub-band had much higher noise, which resulted in the smaller range of searchable \HI~velocities in A and D, as compared to B and C ({$-$1000} \kms~to 10,500 \kms~and {$-$1000} \kms~to 11,500 \kms, respectively). Additionally, the noise in A and D fields increases significantly at velocities beyond 7000 \kms~due to a low frequency filter used to damp RFI prevalent outside of the protected 1420 MHz band for commensal partners.

\subsection{Unusable Velocities}

There are some velocity ranges in this survey which we cannot easily search for \HI~lines. In the velocity range 8400-8800 \kms~RFI by the L3 GPS around 1381 MHz make it unusable. At 4500-5500 \kms, there is an increase in the noise level in the B and C fields caused by an intermodulation product (intermod) at 1396 MHz that is on for 50\% of the time, but in the B and C fields only. It is caused by the introduction of two or more unwanted signals at integer sums of frequencies. Furthermore, extragalactic \HI~signals might be masked by Galactic \HI~(\vhel ~$\approx~\pm$ 200 \kms), and by High Velocity Clouds (HVCs), which can be found with velocities between $v_{LG}\sim~\pm$500 \kms (\citealt{Putman02}), though these only appear occasionally in our field and they can be distinguished from galaxies with closer examination. 

\section{Search methods and \HI~parametrization} \label{sec:search}

The initial source list for the shallow phase of ALFAZOA was assembled separately for each region of the survey. After the regions were Hanning smoothed and spectral cubes were made for the completed survey area, each region was searched by two or three authors, who visually inspected the cubes using the visualization package KARMA (\citealt{Gooch96}) and identified possible sources, compiling independent lists of candidates. Although other HI surveys use automatic source detection software to produce their source lists, we did not, as the source finders at the time we tested performance (2014) produced an unmanageable number of false detections in the more complicated ZOA, due to increased continuum emission, and the human eye was better for finding galaxies and rejecting spurious signals. These lists were compared and matched in position and velocity using the astronomical coordinate comparison tool, TOPCAT (\citealt{Taylor05}), to deal with redundancies (the position is matched to 4', and the velocity to 300 \kms). The list, including all candidates, was passed on to another author, who adjudicated all candidates by re-examining every source, either accepting it into the final catalog or rejecting it as a false detection. 

Although we did not consider particular selection criteria for the initial visual search, the resulting \HI~sources in the final catalog all have the following characteristics: (1) their peak is at or above the 3 $\sigma$ rms noise, (2) they are at a velocity where we can be confident that it is separated from Galactic velocities (where \vhel ~$>$ 100 \kms), and (3) they are not in the middle of well known regions of RFI.

Once the final list was made of all sources in the survey, the \HI~sources were parametrized using the program MBSPECT of the software package MIRIAD (\citealt{Sault95}), in the same way described in \cite{Henning10}. Moment zero maps were made of the \HI~column density distribution of each source, and were used to determine their angular extent and to classify them as extended or unresolved; only about 9\% were clearly extended. The bounding box used to measure the parameters of the source was adjusted according to the size of the source. MBSPECT found the centroid position of the source, and used the weighted sum of the emission along the spectral axis to create one-dimensional profiles of the \HI~spectrum, such as those seen in Figure \ref{profile}. The integrated line flux (\flux) was calculated by subtracting the baseline and integrating across the channels previously determined to have \HI~emission. For all the sources in the survey, 66\% of them have a subtracted baseline of order 1, 23\% of order 2 and 11\% are order 3.  There is a single source which used order 4. The baseline was chosen on a case-by-case basis by examining the shape of the baseline in each individual spectrum, and finding an order which best represented it.  Each subtracted baseline is shown on each spectrum in Figure \ref{profile}.The heliocentric velocity (\vhel) for each source is measured as the middle point of the velocities at 50\% of peak flux density. The line widths at 50\% and 20\% of peak flux density (\wfifty~and \wtwenty, respectively) were measured using the width-maximizing algorithm applied by MBSPECT. 

\section{The ALFAZOA Shallow Survey Catalog} \label{sec:over}

\begin{figure*}[htbp]
\includegraphics[width=1.0\textwidth]{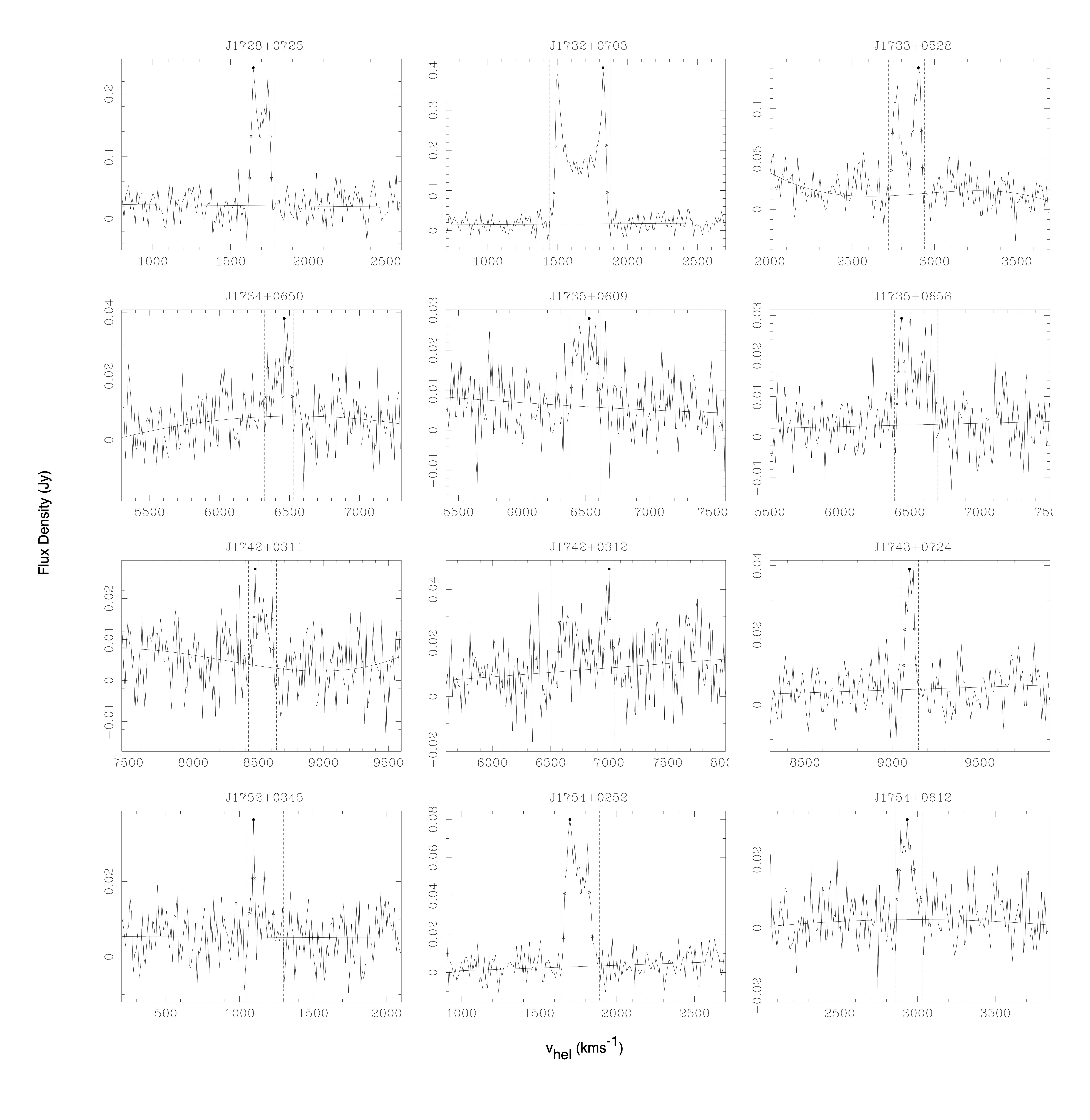}
\caption{Examples of \HI~spectra for galaxies detected in the ALFAZOA Shallow survey, separated by field. The vertical dashed lines represent the boundaries of the measured profile, the solid line represents the baseline to be subtracted from the spectrum, the solid black dot is the location of the peak flux, the circles are the width-maximizer widths, \wfifty~and \wtwenty, and the x's are the width-minimizer widths, which are not used in this paper. The complete figure set (34 pages) is available electronically.}
\label{profile}
\end{figure*}

\begin{figure}[tb]
\begin{center}
\includegraphics[width=0.45\textwidth]{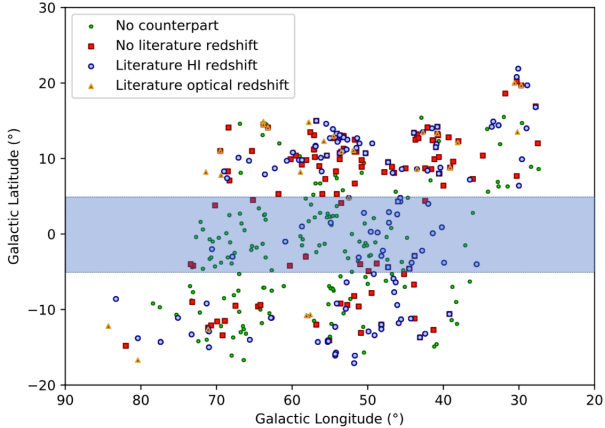}
\caption{A map of the sky distribution of all \HI~detections from the ALFAZOA Shallow survey, in Galactic coordinates. Green circles represent sources without a counterpart, Red squares are possible counterparts without recorded redshift information, blue circles represent all counterparts with previous \HI~measurements, including HIZOA and HIPASS sources, and yellow triangles represent counterparts with only optical redshift information. The region between $\pm 5$\degree~of $b$=0\degree~is marked out in light blue. Note how most counterparts (excluding blue \HI~detections) are from outside this region.}
\label{cmap}
\end{center}
\end{figure}

The ALFAZOA Shallow Survey detected 403 galaxies, with velocities between 281 {\kms} $\leq$ \vhel $\leq$ 11840 \kms, and profile widths between 16 \kms $\leq$ \wfifty $\leq$ 627 \kms. They are located both inside the complete survey region ($\lvert b \lvert $ $\leq$ 10\degree~and 30\degree~$\leq$ l $\leq$ 75\degree) and in the regions of limited sensitivity (see Figures \ref{coverage} and \ref{cmap}; the first shows the regions observed, and the second the positions of all ALFAZOA Shallow detections).

Table 2 is an example page of the full catalog, which is available electronically, and it presents the measured and derived parameters for the galaxies in the columns described below:

${Column ~(1)}$ -- ALFAZOA source name;

${Columns ~(2)~and ~(3)}$ -- Equatorial coordinates (J2000) of the fitted centroid position;

${Columns~(4) ~and ~(5)}$ -- Galactic coordinates of the fitted centroid position;

${Column ~(6)}$ -- Integrated \HI~flux and its uncertainty. Fluxes measured in the A and D fields are indicated with an asterisk. These values do not include the 4\% uncertainty due to the scaling described in section \ref{sec:ad};

${Column ~(7)}$ -- Heliocentric velocity and associated error;

${Columns ~(8) ~and ~(9)}$ -- Velocity width of the profile at 50\% and 20\% of the peak flux density, respectively, and associated uncertainty;

${Column ~(10)}$ -- Distance to the galaxy in the Local Group reference frame (\citealt{Courteau99}). The corrected velocity is calculated as:
\begin{equation}
v_{LG}=v_{hel}+300~\sin l ~\cos b
\end{equation}
 Hubble's law for the cosmological redshift distance was used, with a Hubble constant $H_0$~=~70;
 
 \begin{longrotatetable}
\begin{deluxetable*}{ccccccccccc}

\tablecaption{Properties of ALFAZOA Galaxies}
\tablewidth{700pt}
\tablecomments{This table is published in its entirety in the machine-readable format}

\tablehead{
\colhead{ALFAZOA} & \colhead{R.A.} & 
\colhead{Dec.} & \colhead{l} & 
\colhead{b} & \colhead{flux} & 
\colhead{v$_{hel}$} & \colhead{w$_{50}$} & 
\colhead{w$_{20}$} & \colhead{D$_{LG}$} & 
\colhead{$\rm{log}(\frac{M_{HI}}{M_{\odot}})$} \\
\colhead{} & \colhead{(HH:MM:SS)} & 
\colhead{(DD:MM:SS)} & \colhead{(\degree)} & \colhead{(\degree)}& 
\colhead{(Jy \kms)} & \colhead{(\kms)} & 
\colhead{(\kms)} & \colhead{(\kms)} & \colhead{(Mpc)} & \colhead{}
} 

\startdata
J1728+0725\footnote[1]{Source might be two galaxies}$^e$& 17:28:12 & 07:25:00 & 30.1 & 21.9 & 22.1* $\pm$ 4.0 & 1694 $\pm$ 4 & 122 $\pm$ 9 & 144 $\pm$ 13 & 28.3 & 9.6 \\
J1732+0703$^e$ & 17:32:24 & 07:03:25 & 30.3 & 20.8 & 76.6* $\pm$ 6.0 & 1664 $\pm$ 2 & 364 $\pm$ 5 & 382 $\pm$ 7 & 25.1 & 10.1 \\
J1733+0528$^e$ & 17:33:56 & 05:28:12 & 28.9 & 19.7 & 12.8* $\pm$ 2.4 & 2832 $\pm$ 4 & 176 $\pm$ 8 & 190 $\pm$ 11 & 38.8 & 9.7 \\
J1734+0650 & 17:34:23 & 06:50:51 & 30.3 & 20.2 & 2.0* $\pm$ 0.9 & 6427 $\pm$ 7 & 170 $\pm$ 15 & 183 $\pm$ 22 & 91.1 & 9.6 \\
J1735+0609 & 17:35:19 & 06:09:32 & 29.7 & 19.7 & 2.9* $\pm$ 1.1 & 6500 $\pm$ 8 & 211 $\pm$ 17 & 226 $\pm$ 25 & 90.1 & 9.8 \\
J1735+0658 & 17:35:50 & 06:58:01 & 30.6 & 20.0 & 3.7* $\pm$ 1.0 & 6539 $\pm$ 10 & 243 $\pm$ 20 & 274 $\pm$ 30 & 92.0 & 9.9 \\
J1742+0311 & 17:42:05 & 03:11:49 & 27.8 & 16.9 & 1.9* $\pm$ 1.0 & 8539 $\pm$ 14 & 145 $\pm$ 28 & 178 $\pm$ 42 & 121.2 & 9.8 \\
J1742+0312 & 17:42:14 & 03:12:19 & 27.8 & 16.8 & 4.7* $\pm$ 1.9 & 6791 $\pm$ 15 & 434 $\pm$ 30 & 470 $\pm$ 44 & 96.2 & 10.1 \\
J1743+0724 & 17:43:01 & 07:24:39 & 31.8 & 18.6 & 1.7* $\pm$ 0.6 & 9100 $\pm$ 5 & 56 $\pm$ 10 & 70 $\pm$ 15 & 131.6 & 9.8 \\
J1752+0345 & 17:52:59 & 03:45:14 & 29.6 & 14.7 & 1.0* $\pm$ 0.7 & 1130 $\pm$ 18 & 81 $\pm$ 37 & 166 $\pm$ 55 & 18.4 & 7.9 \\
J1754+0252 & 17:54:35 & 02:52:47 & 29.0 & 14.0 & 9.5* $\pm$ 1.3 & 1744 $\pm$ 4 & 158 $\pm$ 8 & 186 $\pm$ 12 & 24.4 & 9.1 \\
J1754+0612 & 17:54:20 & 06:12:15 & 32.0 & 15.5 & 2.4* $\pm$ 1.0 & 2925 $\pm$ 13 & 109 $\pm$ 27 & 155 $\pm$ 40 & 39.5 & 8.9 \\
J1758+0041 & 17:58:59 & 00:41:13 & 27.5 & 12.0 & 3.8* $\pm$ 1.2 & 3898 $\pm$ 12 & 64 $\pm$ 23 & 142 $\pm$ 35 & 58.2 & 9.5 \\
J1758+0343 & 17:58:12: & 03:43:23: & 30.2 & 13.5 & $\cdots$* & 327: & $\cdots$ & $\cdots$ & $\cdots$ & $\cdots$ \\
J1759+0617$^e$ & 17:59:29 & 06:17:00 & 32.7 & 14.4 & 34.1* $\pm$ 4.0 & 1808 $\pm$ 5 & 243 $\pm$ 10 & 281 $\pm$ 16 & 24.8 & 9.7 \\
J1759+0708$^1$$^e$ & 17:59:08 & 07:08:01 & 33.4 & 14.9 & 20.5* $\pm$ 2.8 & 1889 $\pm$ 4 & 199 $\pm$ 9 & 224 $\pm$ 13 & 24.4 & 9.5 \\
J1801+0657$^1$$^e$& 18:01:53 & 06:57:40 & 33.6 & 14.2 & 82.0* $\pm$ 6.5 & 1956 $\pm$ 3 & 315 $\pm$ 6 & 339 $\pm$ 9 & 27.8 & 10.2 \\
J1803+0722 & 18:03:49 & 07:22:03 & 34.2 & 13.9 & 1.4* $\pm$ 0.6 & 1805 $\pm$ 12 & 69 $\pm$ 25 & 126 $\pm$ 37 & 26.1 & 8.4 \\
J1808+0459 & 18:08:17 & 04:59:43 & 32.5 & 11.9 & 0.8* $\pm$ 0.5 & 6551 $\pm$ 6 & 24 $\pm$ 11 & 37 $\pm$ 17 & 96.5 & 9.3 \\
J1810-0105 & 18:10:46 & -01:05:15 & 27.4 & 8.6 & 3.1* $\pm$ 1.2 & 2108 $\pm$ 7 & 139 $\pm$ 15 & 153 $\pm$ 22 & 27.9 & 8.8 \\
J1810+0135$^1$$^e$ & 18:10:24 & 01:35:12 & 29.6 & 9.9 & 8.2* $\pm$ 2.7 & 1802 $\pm$ 10 & 155 $\pm$ 20 & 195 $\pm$ 30 & 30.0 & 9.2 \\
J1811-0004 & 18:11:37 & -00:04:06 & 28.4 & 8.9 & 2.8* $\pm$ 1.4 & 1687 $\pm$ 8 & 120 $\pm$ 16 & 134 $\pm$ 24 & 24.6 & 8.6 \\
\enddata

\label{galaxies}
\end{deluxetable*}
\end{longrotatetable}

${Column ~(11)}$ -- Logarithm of the total \HI~mass, calculated using:
\begin {equation}
M_{\HI}=2.36\times10^5~D_{LG}^2~F_{\HI}
\end{equation}

\begin{figure}[b]
\begin{center}
\includegraphics[width=0.45\textwidth]{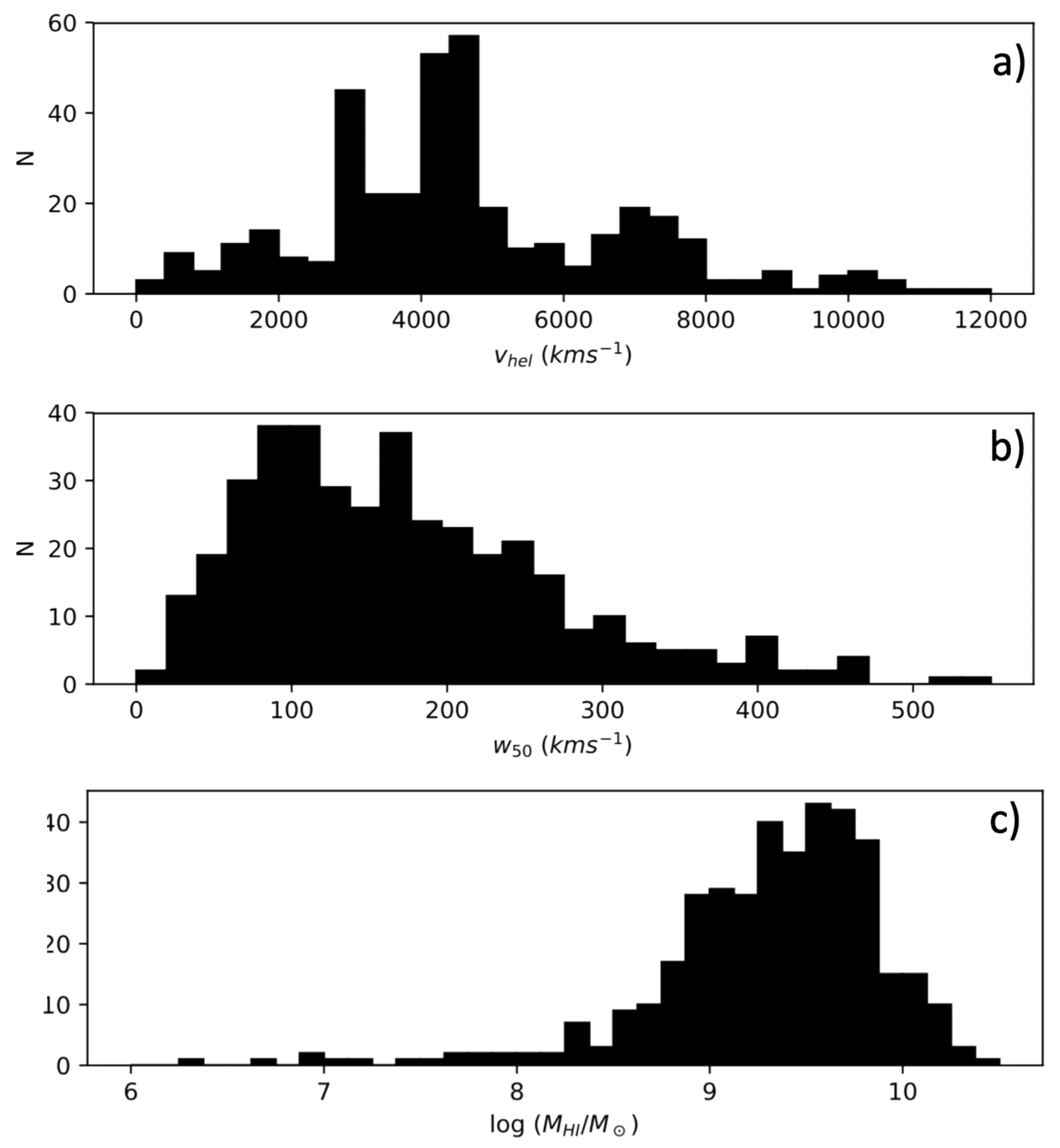}
\caption{Histograms of the distribution of three HI~parameters of the ALFAZOA Shallow Survey galaxies. From top to bottom, they show (a) heliocentric velocity, (b) velocity width at 50\% of peak flux density, and (c) \HI~mass distribution.}
\label{histogram}
\end{center}
\end{figure}

The uncertainties on \flux, \vhel, \wfifty, and \wtwenty~listed in the table were calculated using the formalism of \cite{Koribalski04}, and do not take into account baseline fitting errors. Because the uncertainties on \distance ~and \mass~rely heavily on cosmological assumptions, their uncertainties are not included.

For thirteen sources that lie on the edges of the survey region the fluxes could not be completely recovered. They are distinguished in the table with a colon after their position and velocity, since they could not be fitted in the way described in Section \ref{sec:search}. Their derived parameters are not calculated and are indicated with ellipses.

The \HI~profiles for 12 out of the 403 galaxies detected in the ALFAZOA Shallow Survey are shown in Figure \ref{profile}. The noise in regions A and D is higher by 1.6 mJy than the noise measured in the B and C. Largely due to the difference in noise, the number of detected galaxies in A and D regions is almost half the number found in the B and C regions. Additionally, the the noise beyond \vhel=7000 \kms~is even higher, and this velocity interval accounts for about 20\% of the total galaxy detections in B and C. The rest of the spectral profiles appear in the online version of the journal.

Panel $a$ of Figure \ref{histogram} shows the heliocentric velocity distribution. The decrease in the number of distant galaxies corresponds to the drop-off in sensitivity, as well as to the increased noise above \vhel=7000 \kms~for half of the survey. We can also observe, despite the galaxies being averaged over unrelated LSS in our survey region, overdensities at $\sim$ 3000 km/s and 4400 km/s, the first of which is a new structure and the second which appears to connect to the Pegasus overdensity, from the {\cite{Erdogdu06}} nomenclature. The LSS is discussed further in Section \ref{sec:LSS}. Panel $b$ shows the distribution of the \HI~line velocity widths for the galaxies. The distribution is non-Gaussian, with a median value of 158 \kms~and a mean of 174 \kms. This is comparable to the Parkes' HIZOA survey (\citealt{Stavely16}), a survey of 883 galaxies with a median of 147 \kms~and a mean value of 163 \kms. Figure \ref{histogram}-c shows the \HI~mass distribution, which ranges from \mass=$10^{6.3} M_{\odot}$ to $10^{10.5} M_{\odot}$. The median mass is $10^{9.4} M_{\odot}$, and the mean is $10^{9.6} M_{\odot}$.


\section{Counterparts} \label{sec:counter}

A list of counterparts of the \HI~detections was created by searching the NASA/IPAC Extragalactic Database (NED)\footnote{The NASA/IPAC Extragalactic Database (NED) is operated by the Jet Propulsion Laboratory, California Institute of Technology, under contract with the National Aeronautics and Space Administration.} for extragalactic sources within a radius of 2 arcmin of each galaxy's position. The matches were considered possible counterparts if they were classified in NED as galaxies, and for those with redshift values, if their recorded velocity was within 300 \kms~of the ALFAZOA source. Table 3 is an example page of the data we collected on ALFAZOA sources with counterparts in the literature (the full table can be read electronically) with the following columns:

${Column ~(1)}$ -- ALFAZOA source name;

${Columns~(2) ~and ~(3)}$ -- Galactic coordinates of the fitted \HI~centroid position;

${Column ~(4)}$ -- Foreground extinction A$_{B}$ calculated by using the E(B-V) values from the IRAS DIRBE maps of \cite{Schlafly11}, and adopting $R_B$=4.14, in magnitudes;

${Column ~(5)}$ -- The counterpart galaxy identifier, going by the foremost name used in NED;

${Columns ~(6)}$ -- Identification of the wavelengths of previous observations as found in NED: "I" for infrared, "O" for optical, and "H" for an \HI;

${Columns ~(7)}$ -- Distance between the positions of the \HI~centroid and the counterpart;

${Column ~(8)}$ -- Separation in velocity between our measured \HI~\vhel~and the recorded velocity in the literature (where available); 

There is sometimes more than one possible match for each galaxy. An attempt to narrow down the possible matches is only made in the cases when the velocity data is available, in which case galaxies with velocities that are separated by more than 300 \kms~from our source are excluded. Otherwise, all possible counterparts are listed. In the cases where the source was extended or the velocity width was larger than 400 \kms, both the search radius and the cutoff for the velocity difference were increased to scale with the size and velocity width of the source. 

\subsection{Results}

\begin{figure}[b]
\begin{center}
\includegraphics[width=0.45\textwidth]{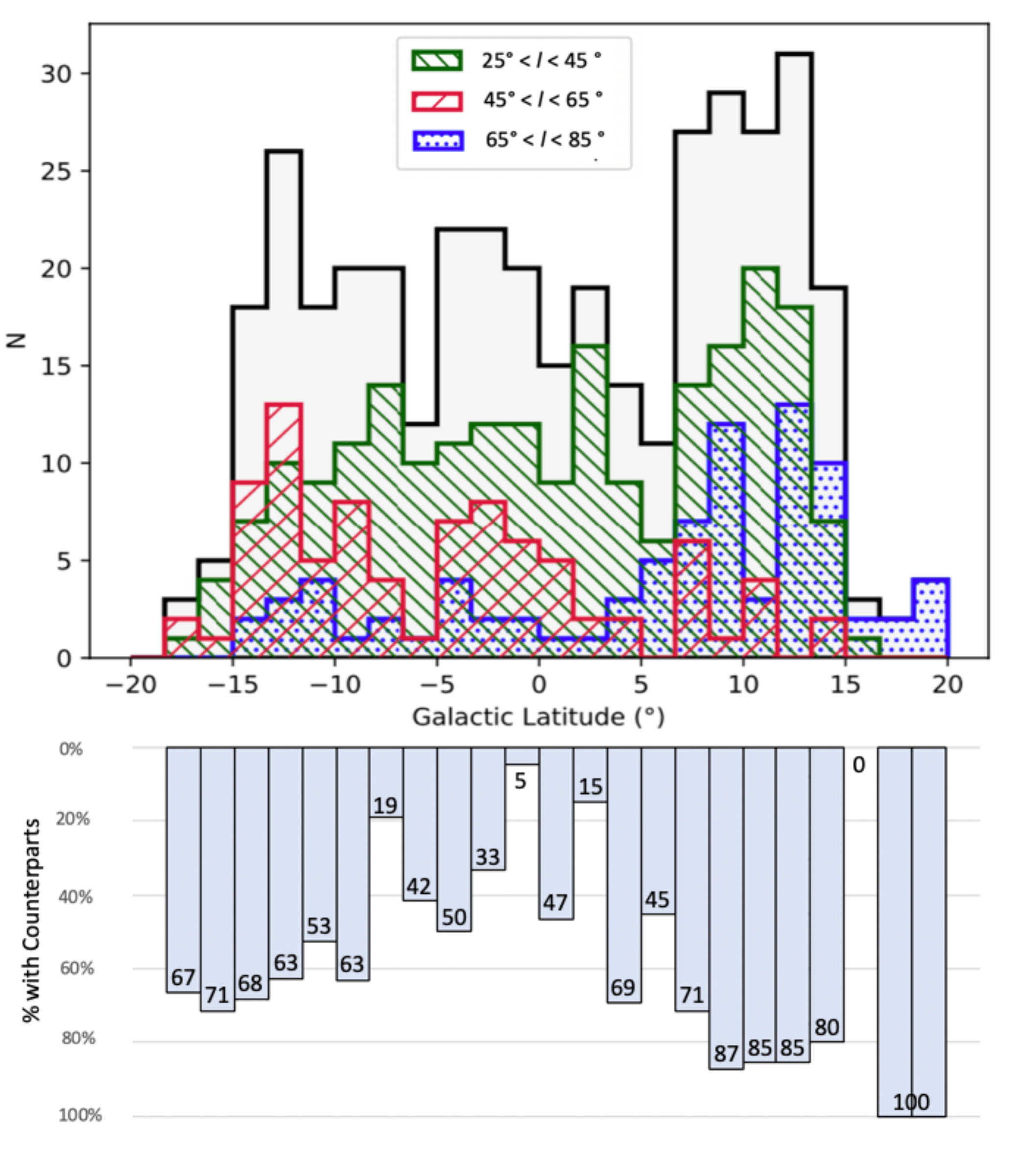}
\caption{Top: a histogram of the galactic latitude distribution of all ALFAZOA Shallow detections. The colored, patterned regions represent different intervals of galactic longitude of the sky: green for 25\degree $\leq l \leq 45$\degree, red for 45\degree $\leq l \leq$ 65\degree~and blue for 65\degree $\leq l \leq$ 85\degree. Bottom: a histogram of the percentage of galaxies with counterparts for each galactic latitude bin. As expected, the fewest counterparts are found in the region between $b=\pm$5\degree. The increase in detections between 8\degree~$\leq l \leq$ 14\degree~corresponds to an overdensity in that region.}
\label{chist}
\end{center}
\end{figure}

\begin{figure*}[htbp]
\begin{center}
\includegraphics[width=0.9\textwidth]{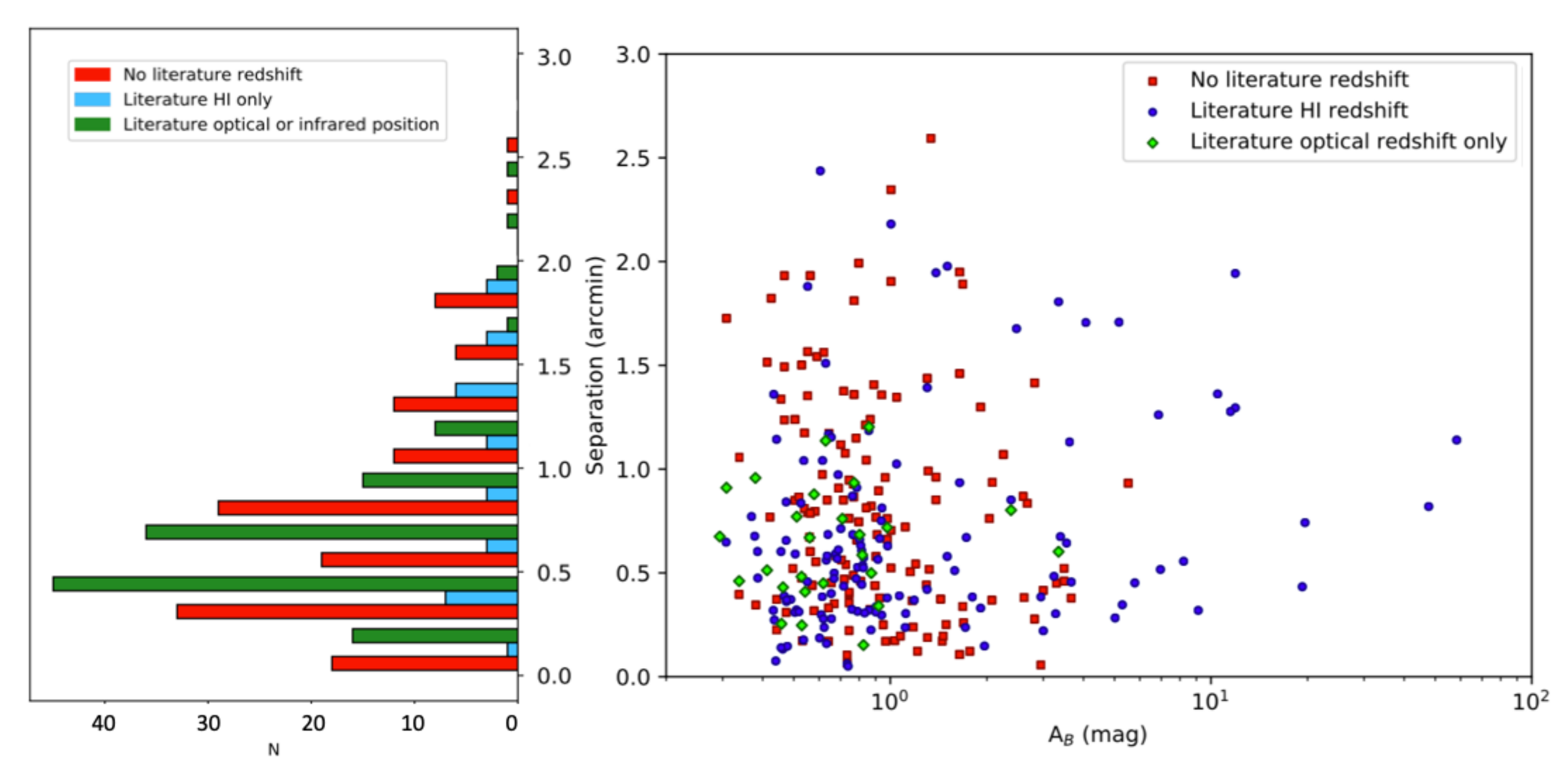}
\caption{[Left] Histogram of the distribution of separation distances between positions of candidate counterpart galaxies and ALFAZOA sources. The red represents counterparts with no prior velocity information, the green represents counterparts with velocity information as well as an optical or infrared position, and blue those with only previous \HI~information. For comparison, half the FWHM of the Arecibo telescope beam is 1.7 arcmin. [Right] The separation distance between the positions of ALFAZOA sources and their counterparts (in arcmin) as a function of foreground extinction (A$_B$), in magnitudes. Blue circles represent all counterparts with \HI~measurements, red squares represent counterparts without redshift information and green diamonds represent those with only optical redshift measurements. The majority of counterparts in high-extinction areas (A$_B$ $>$ 3 mag) are other \HI~detections . }
\label{sep}
\end{center}
\end{figure*}

Because the survey focuses on the ZOA, at least 168 of the galaxies were detected in this survey for the first time. Most of the new galaxies are in regions of high extinction close to the Galactic plane, and the likelihood of an \HI~source having a counterpart depends also on its flux and velocity width (e.g. \citealt{Stavely16}). This is illustrated in Figures \ref{cmap}, \ref{chist}, \ref{sep} and \ref{velmass}. Out of the 403 sources, 235 have at least one possible counterpart (58\%). With only one exception (J2006+3505), all galaxies with an integrated flux above 15 Jy \kms~have a counterpart in the literature. Similarly, 84\% of galaxies with a velocity width higher than 300 \kms~have counterparts.

Of the 235 sources with possible counterparts, 144 of them (61\%) have previous spectroscopic redshift measurements, mostly from 21-cm \HI~(89\%). This includes 27 detections (11\% of all counterparts) which only have prior \HI~information (9 from HIPASS, 15 from HIZOA and 3 from ADBS), 65 sources with counterparts that have optical redshifts (28\%), and 85 (36\%) have no redshift information in the literature. Out of all the sources with possible counterparts, 176 of them (74\%) have at least one counterpart from 2MASS. The average spatial separation between our detections and the literature galaxies is 0.7'. Excluding the 47 detections with more than one possible counterpart, the average separation distance is 0.6'.

\startlongtable
\begin{deluxetable*}{ccccccccc}
\tabletypesize{\scriptsize}

\tablecaption{Possible ALFAZOA Counterparts \label{counterparts}}
\tablewidth{700pt}
\tablecomments{This table is published in its entirety in the machine-readable format}

\tablehead{ \\
\colhead{ALFAZOA} & 
\colhead{l} & 
\colhead{b} & 
\colhead{A$_B$} & 
\colhead{Name} & 
\colhead{IOH}  & 
\colhead{Sep} & 
\colhead{$\delta$v}\\
\colhead{} & 
\colhead{(\degree)} & 
\colhead{(\degree)} & 
\colhead{mag} & 
\colhead{} & 
\colhead{} &
\colhead{(arcmin)} & 
\colhead{(\kms)}
} 

\startdata
J1728+0725 & 30.1 & 21.9 & 0.37 & UGC 10862 & I O H & 0.77 & 4 \\
J1732+0703 & 30.3 & 20.8 & 0.44 & NGC 6384 & I O H & 0.08 & -1 \\
J1733+0528 & 28.9 & 19.7 & 0.55 & UGC 10901 & - - H & 0.46 & -2 \\
J1734+0650 & 30.3 & 20.2 & 0.50 & 2MASX J17342223+0651434 & I - - & 0.85 & \\
& & & & 2MASX J17341908+0650483 & I - - & 1.24 & \\
J1735+0609 & 29.7 & 19.7 & 0.41 & CGCG 055-010 & I O - & 0.51 & -20 \\
& & & & 2MASX J17351335+0609098 & I - - & 1.52 & \\
J1735+0658 & 30.6 & 20.0 & 0.46 & CGCG 055-011 & I O - & 0.25 & 55 \\
J1742+0311 & 27.8 & 16.9 & 0.92 & 2MASX J17420123+0312013 & I - - & 0.90 & \\
J1742+0312 & 27.8 & 16.8 & 0.94 & UGC 10943 & I O H & 0.81 & -2 \\
J1743+0724 & 31.8 & 18.6 & 0.54 & 2MASX J17425821+0725174 & I - - & 0.81 & \\
J1754+0252 & 29.0 & 14.0 & 0.69 & UGC 11030 & I O H & 0.61 & -12 \\
\enddata
\label{counterparts}
\end{deluxetable*} 

The right side of Figure \ref{sep} shows the separation between the ALFAZOA sources and their counterparts as a function of foreground extinction, illustrating how the galaxies without redshift information make up most of the counterparts with separations above 1.0 arcmin. Conversely, all galaxies with a literature counterpart that has only optical redshifts (the green diamonds) have low extinction and small separations (below 1.3 arcmin), while those with literature \HI~or other spectroscopic redshift exist at the full range of separation, but are the only ones found with high extinction.

Considering only the $b \leq \pm$ 10\degree~region covered by ALFAZOA Shallow, 66 (28\%) out of the 237 detected galaxies have 2MASS counterparts. The far lower percentage of NIR counterparts near the Galactic plane is expected due to stellar crowding, and this region has fewer counterparts in any wavelength (Figure \ref{cmap}).

\subsection{Counterpart Characteristics}

\begin{figure}[hbp]
\begin{center}
\includegraphics[width=0.45\textwidth]{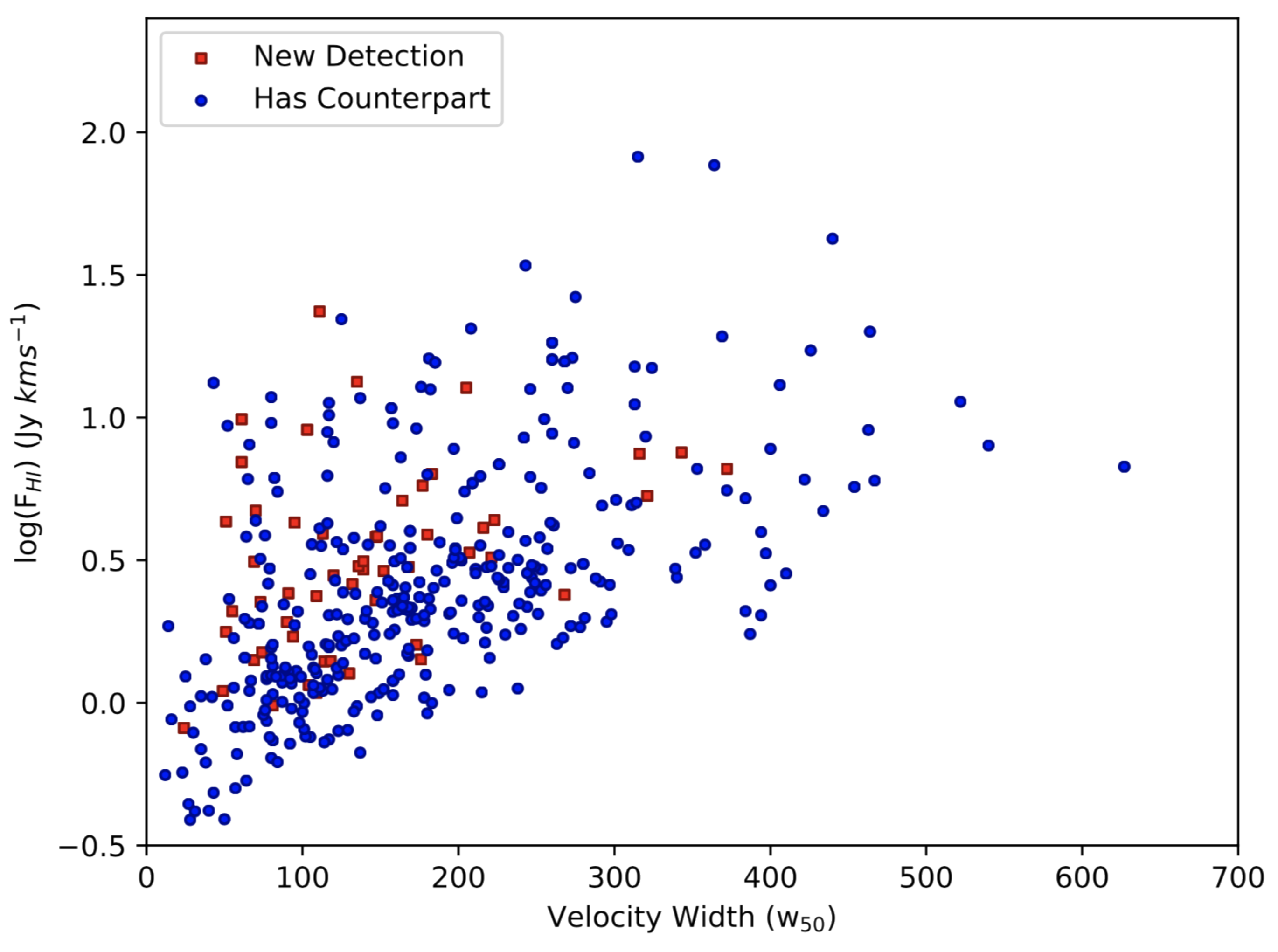}
\caption{The integrated \HI~line flux \flux~(in Jy \kms) as a function of \wfifty velocity (in \kms). The blue circles have possible counterparts, and the red squares are new detections.}
\label{velmass}
\end{center}
\end{figure}

As was previously discussed and illustrated in Figures \ref{cmap}, \ref{chist} and \ref{velmass}, for the \HI~sources the probability of finding a possible counterpart decreases with foreground extinction and increases with their \HI~flux, although there are many sources in areas with relatively low $A_B$ that have no counterpart, as seen in Figure \ref{cmap}. However, there are other factors to be considered for finding possible counterparts of galaxies seen in a blind \HI~survey such as ALFAZOA. 

\HI~21-cm observations are sensitive to gas-rich galaxies which can be of modest optical and NIR luminosity and/or surface brightness, whereas optical and NIR observations are more sensitive for detecting luminous high-surface brightness spirals and active galaxies (\citealt{Haynes11}). However, inside of the ZOA it is difficult to obtain optical detections, due to the relatively high extinction at these wavelengths, and so detecting the low luminosity, low surface brightness galaxies is unlikely. Therefore, low \HI~mass sources are less likely to have counterparts, since they have the highest fraction of gas to stellar mass (\citealt{Roberts94}). High \HI~mass sources, on the other hand, as well as galaxies with broad velocity widths typically have more stars, so they are more likely to have a literature counterpart.

Figure \ref{velmass} shows the integrated flux \flux~as a function of the \wfifty~line width of all ALFAZOA sources, color coded for detections that have a counterpart in NED. With one exception, all sources with high-flux (\flux $\geq $15 Jy \kms) have counterparts, and the exception is a galaxy with \wfifty = 111 \kms and in a region of high extinction (A$_B$=12.10). Similarly, all but two of the high velocity width (\wfifty $\geq$ 350 \kms) sources have counterparts, while only one third of detections with \wfifty = 70 \kms~or below have a counterpart. At the lowest fluxes, \flux $\leq$ 1 Jy \kms, only 9 out of 45 galaxies have an optical or NIR counterpart. 

\begin{figure}[htbp]
\begin{center}
\includegraphics[width=0.45\textwidth]{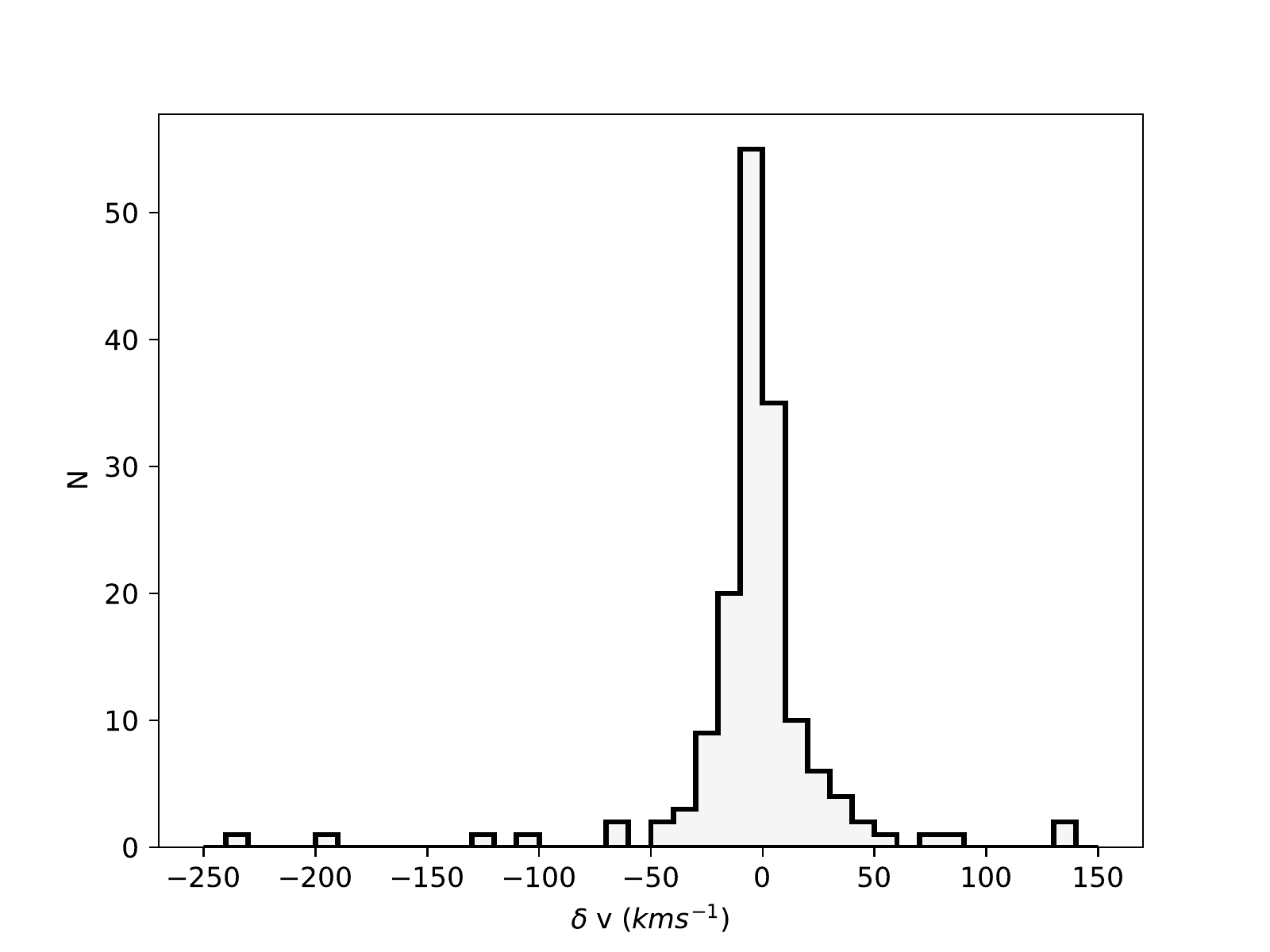}
\caption{Histogram of the distribution of differences $\delta$v between the radial velocities of ALFAZOA sources and their literature counterparts, in \kms.}
\label{exthist}
\end{center}
\end{figure}

When considering the velocity offset between the ALFAZOA Shallow sources and their literature counterparts, the average value of the differences in heliocentric velocity is $-$6.5 \kms, and there does not seem to be any systemic offset in the velocities (Figure \ref{exthist}). The FWHM of the distribution is 86 \kms, which is dominated by the larger uncertainties in the optical velocities of the counterparts compared to the few \kms ~in our HI values. The details of the outliers are described in full in Table 3.

The mean $<$\wfifty$>$ velocity width for detections with a literature counterpart in 2MASS is 213 \kms~compared to 135 \kms~for those without a 2MASS counterpart. The mean \HI~mass for sources with 2MASS counterparts is log $<$ \mass / $M_{\odot}$ $>$ = 9.69 and for those without it is log $<$ \mass / $M_{\odot}$ $>$ = 9.44 $M_{\odot}$. This implies that higher \HI~mass, larger galaxies are more likely to have NIR counterparts.


\section{Survey Performance} \label{sec:selfun}

\begin{figure}[htbp]
\begin{center}
\includegraphics[width=0.45\textwidth]{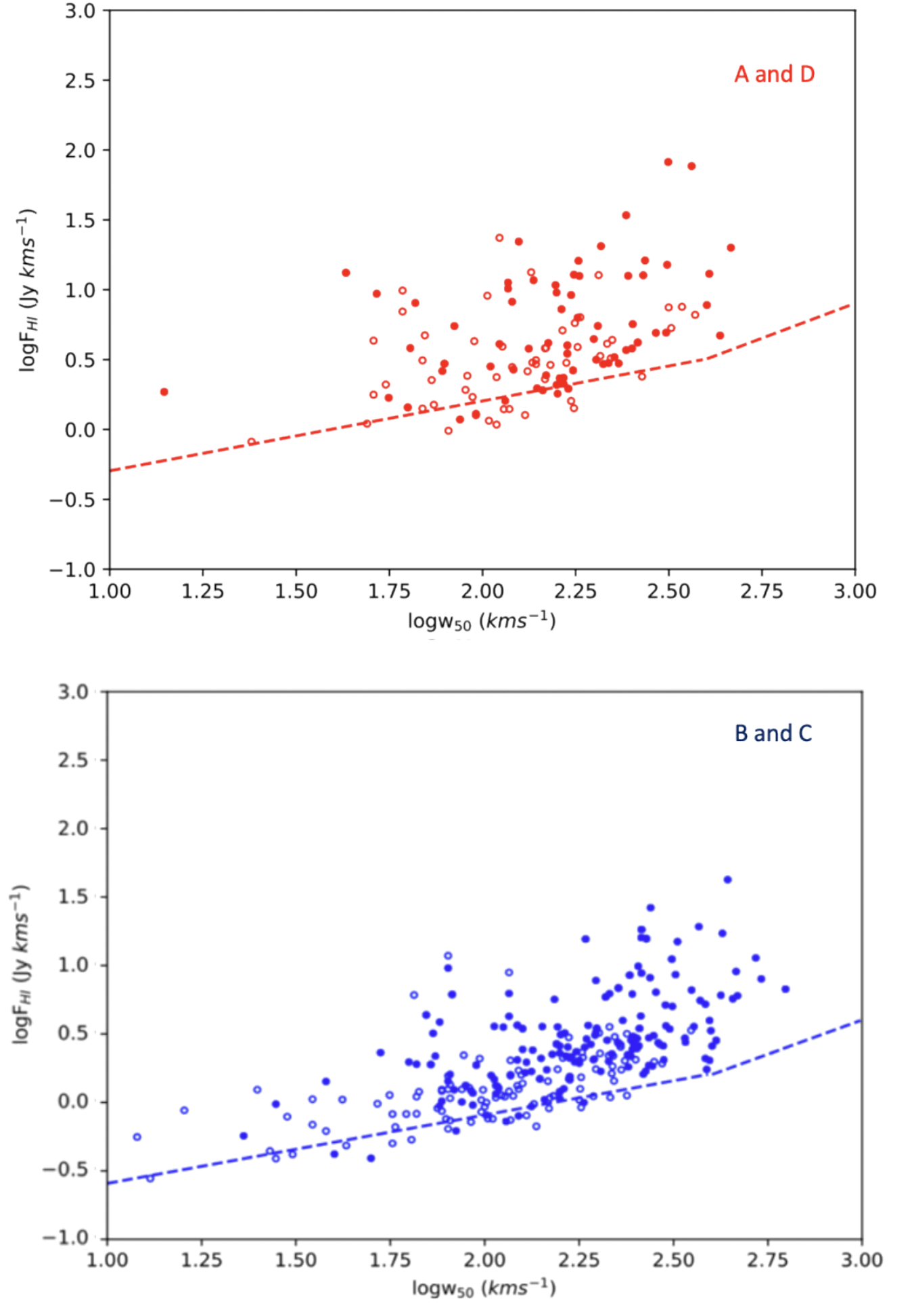}
\caption{Integrated \HI~lie flux \flux~(in Jy \kms) as a function of the \wfifty~line width (in \kms) for all ALFAZOA detections. The top panel shows the A and D field sources and the bottom shows B and C field sources. Open circles represent those without counterparts, and filled circles those with at least one counterpart. The fields have been plotted separately because of their difference in noise level and velocity resolution. The dotted lines shows the S/N = 5 selection function. }
\label{velflux}
\end{center}
\end{figure}

\subsection{Positional Accuracy}

Given that ALFAZOA is Nyquist sampled, the positional accuracy of the survey is well within the 3.4' FWHM of the telescope's primary beam. The left side of Figure \ref{sep} shows the distribution of separation distance from the ALFAZOA source to the possible counterpart. The literature galaxies we can be most confident are true counterparts to our detections are those with corresponding redshift information (green and blue in the histogram). Of these, those with optical or infrared measurements (in green) have positional accuracies an order of magnitude better than our observations, so any offset is likely due to the positional error from ALFAZOA.

The positional accuracy of every ALFAZOA source has a dependency on its signal-to-noise ratio, which \cite{koribalski2004} estimated as the beam size (3.4') divided by the S/N of the detection. This results in a range of positional accuracy uncertainties from 1.7' to 0.2', depending on the source. If we consider the separations from the 126 optical or infrared counterparts mentioned above, we find that the standard deviation of these separations is $\sigma$ = 0.4'.

The effect of intrinsic offsets between the center position of a galaxy's neutral hydrogen and stellar structures should be negligible here, as an offset of one kiloparsec would correspond to a change of only 3.2 arcsec at 4500 \kms, the mean velocity of the counterparts.

\subsection{Selection Function} \label{sec:selfun2}

Because no automatic source-finding algorithm was used to find our detections, there was no predetermined selection function to begin with. To obtain a selection function and compare it to existing blind \HI~surveys, we compare the detected galaxies' integrated line fluxes to their linewidths (see Figure \ref{velflux}). The dotted line is a flux- and linewidth-dependent S/N of 5, defined following \cite{Saintonge07} and \cite{Cortese08} :

\begin {equation}
S/N=(\frac{1000 \times F_{\HI}}{w_{50}}) \times \frac{w^{\frac{1}{2}}}{rms}
\end{equation}
~
where $w$ is either \wfifty~( 2$\times$ $\delta$v) for linewidths less than 400 \kms, or 400/(2 $\times$ $\delta$v) for linewidths of 400 \kms~or greater, where $\delta$v is the velocity resolution of the survey. As with \cite{Henning10}, 400 \kms~marks the velocity width at which typical spectral baseline fluctuations become comparable to the width of the galaxy profile, as seen in Figure \ref{velflux}. A value of $S/N$=5 includes 84\% of sources in the A and D fields and 96\% in the B and C fields.

 \begin{figure}[htbp]
\begin{center}
\includegraphics[width=0.45\textwidth]{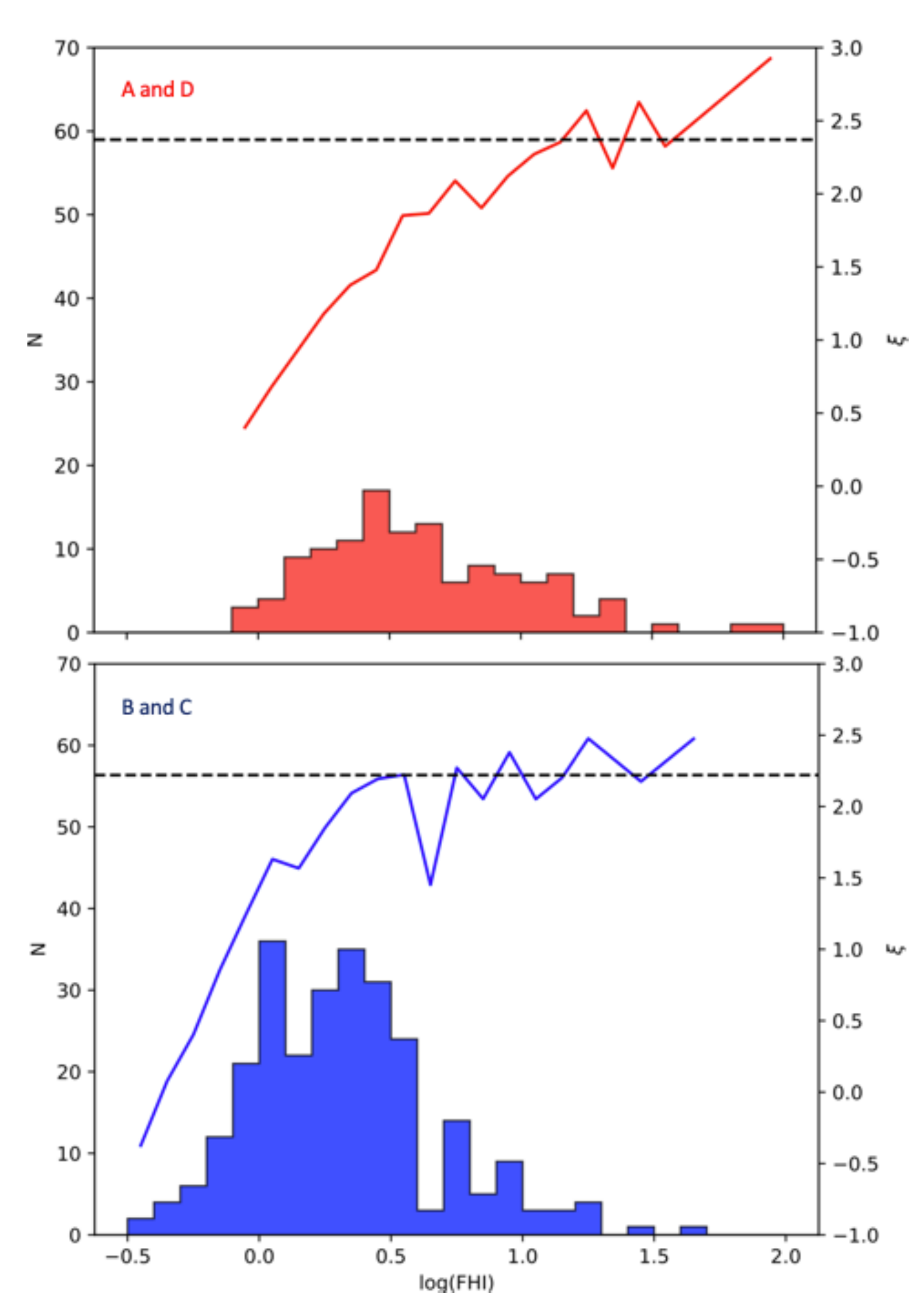}
\caption{Determination of the completeness limit of the survey (see text for details). The top panel is for all A and D field galaxies, the bottom is for B and C. The histograms show the distribution of galaxies with respect to their flux densities, log(\flux). The solid lines are the values of $\xi$, and the dashed line is the best linear fit with a slope of zero.}
\label{complete}
\end{center}
\end{figure}

\subsection{Reliability: B and C fields}

The reliability of the survey, defined as the probability that a detection is real, was determined for the B and C fields through high-sensitivity follow-up observations of 26 sources with varying S/N using the L-band Wide receiver on the Arecibo telescope in 2013. Each source was observed for 180 seconds, using a total power on-off observing mode. Data were taken with the WAPP spectrometer using a velocity resolution of 1.3 \kms. The resulting rms noise was 2.5 mJy at a velocity resolution of 9 \kms. The data were reduced and parametrized with the standard package of IDL programs provided by the Observatory. The results show there are no confirmed sources below S/N=4 and the survey becomes 90\% reliable at S/N=6.6, with each follow-up observation above S/N=7 confirming a detection. Following \cite{McIntyreTh}, we calculated a reliability function which depends on the S/N of the source, according to which the survey was 90\% reliable at a S/N of 6.6, and sources have a 50\% chance of being real at a S/N of 4.4 (for further details, see \cite{McIntyreTh}).

\subsection {Reliability: A and D fields}

We used data from the ongoing ALFAZOA Deep survey to determine the reliability of sources in the A and D fields. The effective integration time for the Deep survey is 5 minutes per beam (compared to 8 sec per beam for the Shallow Survey), with the sky covered by tiles of individual pointings. All of these galaxies, with S/N that ranged from 5 to 13, were confirmed. No reliability estimates are given here due to lack of statistics.

\subsection {Undetected Galaxies}

We looked in the literature for galaxies detected in \HI~which are known to lie within the survey region, but which were not detected by us. We used HyperLEDA\footnote{http://leda.univ-lyon1.fr/} to search for any galaxies which we had not already listed as possible counterparts (inside 2 arcmin radius around each ALFAZOA detection) and found 54 galaxies that fell into this category. 

These 54 galaxies are all associated with faint \HI~sources with profile peaks close to the ALFAZOA survey's noise limit. Out of the 54 galaxies, 32 are in the A and D regions, which had higher noise as described in Section \ref{sec:ad}; the undetected galaxies from A and D had a peak flux density of 20 mJy or below in the literature (corresponding to a peak S/N below 3), with a few exceptions in noisier parts of the cubes, while the undetected galaxies from B and C had a peak flux density of ~12 mJy or less (corresponding to a peak S/N below 3). Half of these undetected galaxies (27 out of 54) can be discerned in the ALFAZOA data with the previous knowledge that they are supposed to be there, although they are still very close to the noise. Seven of these 27 galaxies were considered and rejected during the process of adjudication described in Section \ref{sec:search}, due to being too faint and close to the noise. The remaining 27 have too little signal and are indistinguishable from the noise at the positions and velocities where they are supposed to exist.

\subsection{Completeness}

 \begin{figure}[htbp]
\begin{center}
\includegraphics[width=0.4\textwidth]{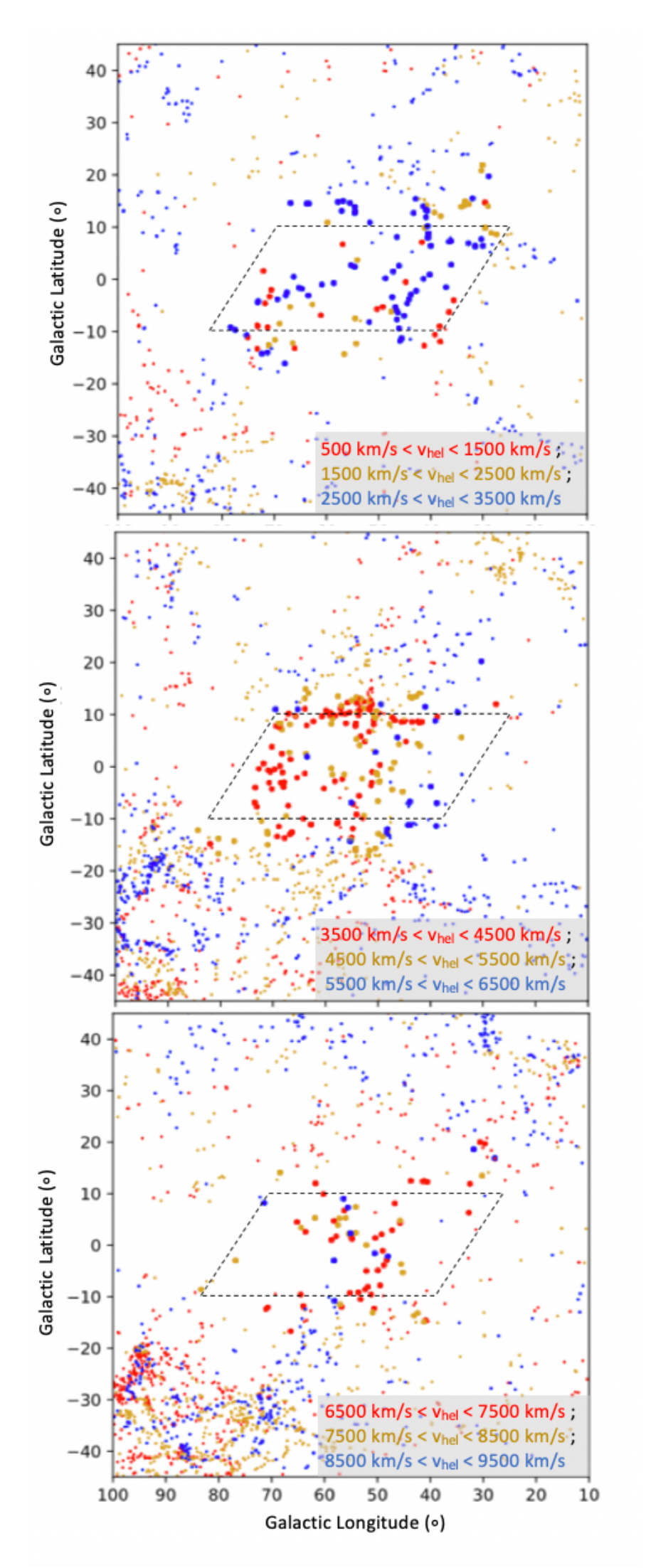}
\caption{Comparison of ALFAZOA Shallow detections to galaxies with redshifts from the HyperLEDA catalog. Three sky projections in Galactic coordinates in redshift intervals of 3000 \kms~width each within 500 $<$ \vhel $<$ 9500 \kms. The ALFAZOA Shallow full sensitivity survey area is marked by the dashed black outline. The most distant slice (higher than 9500 \kms) is not presented here because there were too few detections in this interval (16 in total, see Figure \ref{histogram}). The ALFAZOA detections (larger dots) are combined with galaxies with available redshifts in HyperLEDA up to latitudes of $\lvert b \lvert \leq$ 50\degree to investigate the connection of our detections with known large-scale structures. Each slice has three colors representing a different 1000 \kms~slice of velocity.}
\label{lsslb}
\end{center}
\end{figure}

The most important value in estimating how likely it is for a galaxy to be detected is integrated flux. As illustrated in Figure \ref{complete} smaller fluxes make a galaxy harder to detect. Following this, and the results from previous \HI~surveys (e.g., \citealt{Donley05}), we expect the survey to be mean flux density limited, where the mean flux density is the integrated flux divided by the velocity width at 50\% of peak flux. To estimate the completeness limit, we start by defining a parameter $\xi$ to fit to a histogram of the log(\flux) distribution:

\begin {equation}
\xi=log(F_{\HI}^{3/2}\frac{dN}{dlogF_{\HI}})
\end{equation}

 \begin{figure*}[htbp]
\begin{center}
\includegraphics[width=0.9\textwidth]{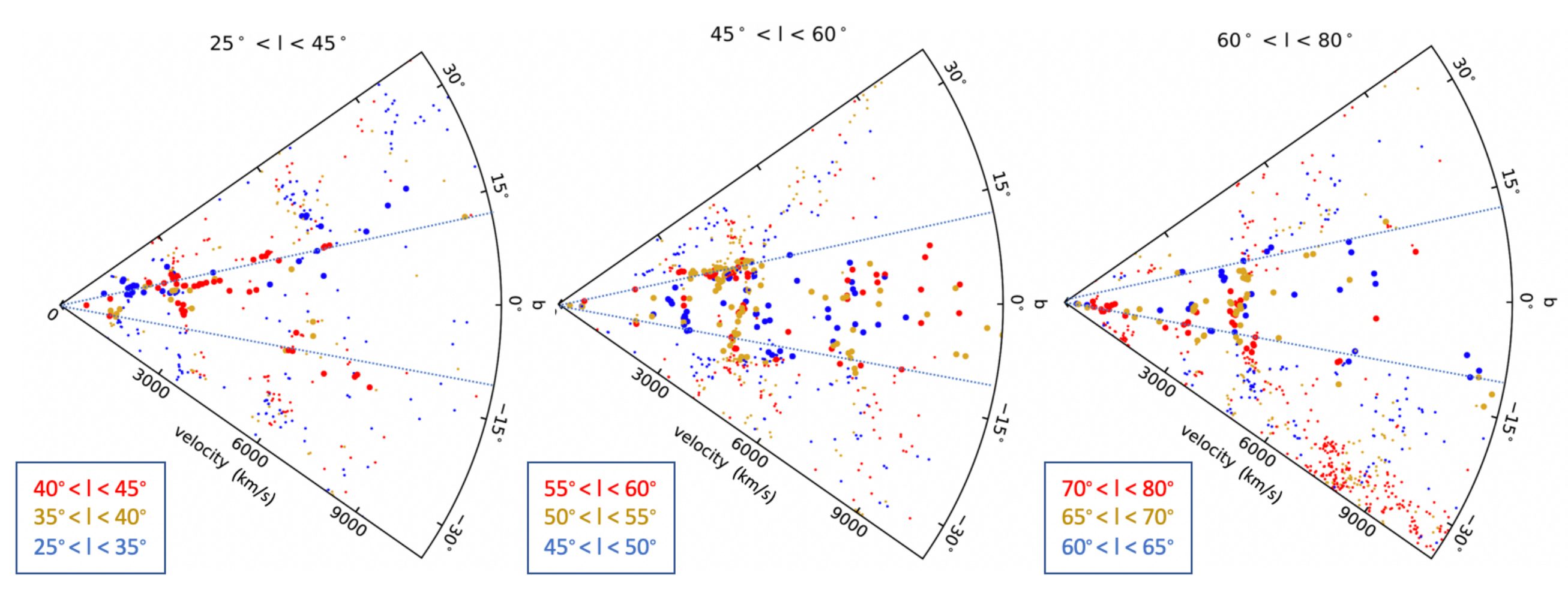}
\caption{Radial velocities of galaxies as a function of Galactic latitude $b$, in nine slices of Galactic longitude $l$ of 5\degree width each, within 25\degree $< l <$ 80\degree. Detections from ALFAZOA Shallow are shown as large circles and HyperLEDA catalog objects as small dots. The latitude slices are color-coded and labeled in each slice. The full sensitivity survey area of each field is shown by the dotted diagonal lines.}
\label{lsspolar}

\includegraphics[width=0.9\textwidth]{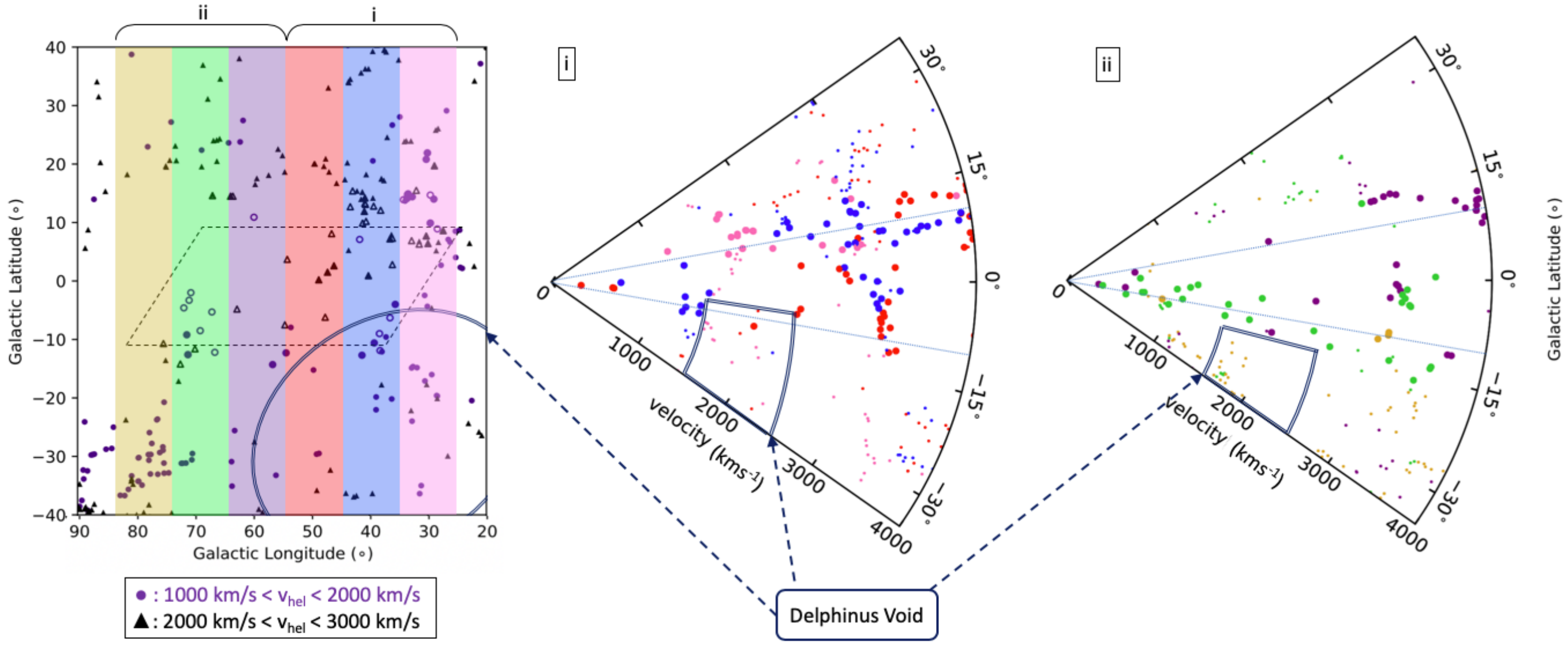}
\caption{(Left) Sky distribution of galaxies in the region with known velocities in the range 1000 \kms $\leq$ \vhel~$\leq$ 3000 \kms, with the full sensitivity ALFAZOA Shallow survey area outlined with a black dashed line. Indigo circles represent galaxies in the 1000-2000 \kms interval, and black triangles at 2000-3000 \kms. The open data points correspond to detections from ALFAZOA, while the filled in ones are from HyperLEDA. (Middle and Right) Wedge plots of the region shown in left projected onto latitude and velocity, with the data points color coded according to the 10\degree~wide colored-in longitude slices on the left. The full resolution area is outlined with grey dotted lines. The velocity ranges from 0 \kms~to 4000 \kms, with an additional 1000 \kms~added at both ends to permit the identification of structures across velocity. The middle plot shows galaxies at 25\degree $\leq l \leq$ 55\degree, corresponding to the pink, blue and red stripes. The plot on the right shows the galaxies at 55\degree $\leq l \leq$ 85\degree, corresponding to the purple, green and yellow stripes. The predicted Delphinus void LSS from \cite{Erdogdu06} is outlined in dark blue and labeled in all three plots.}
\label{lss2000}
\end{center}
\end{figure*}

\bigskip

where dN/d log\flux is the number of detections in each bin of log flux, similar to the method used by \cite{Haynes11} and \cite{McIntyreTh}. A plot of $\xi$ against log \flux~gives a flat distribution for integrated fluxes above the completeness limit, as seen in Figure \ref{complete}. The data are fit from the highest flux to lower flux bins with a linear fit of slope zero, and the completeness limit is determined where the $\chi^2$ of the fit begins to systematically increase. For galaxies in the A and D regions, this happens at log (\flux) = 0.45, where the value of $\xi$ is 2.2$\sigma$ below the fit. A 2.2$\sigma$ event occurs about 1\% of the time in a random distribution, and so the completeness limit, \flux$_{lim}$, is estimated to be \flux$_{lim}$=2.8 Jy \kms. For the B and C fields, the corresponding values are log(\flux) = 0.3 and \flux$_{lim}$=2.0 Jy \kms. 

\begin{figure*}[htb]
\begin{center}
\includegraphics[width=0.9\textwidth]{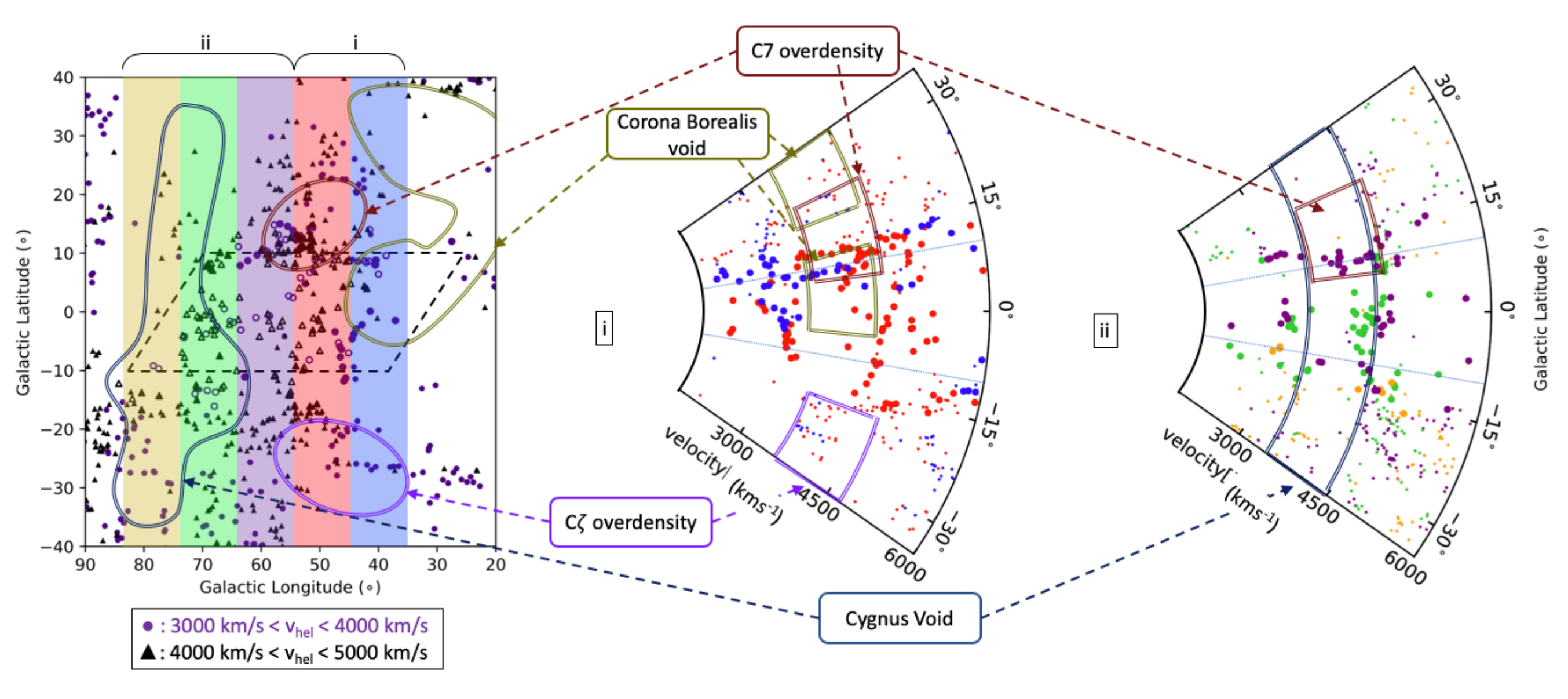}
\caption{Same as in Figure \ref{lss2000}, but with (left) velocities in the range 3000 \kms $\leq$ \vhel~$\leq$ 5000 \kms and (middle and right) velocity ranging from 2000 \kms~to 6000 \kms, with an additional 1000 \kms~added at both ends. Of the four structures shown here, only three intersect the survey area (C7, Corona Borealis and Cygnus). }
\label{lss4000}
\end{center}
\end{figure*}

\section{Large Scale Structure} \label{sec:LSS}

Here we investigate the large-scale distribution of the galaxies detected in this survey, how they relate to known large-scale structures, and examine newly identified structures. To study the relationship with known large-scale structures, we use the \HI~redshifts available from the HyperLEDA (\citealt{Makarov14}), a public data archive, as well as the LSS naming convention and density maps of \cite{Erdogdu06}. Figures \ref{lsslb} and \ref{lsspolar} show the general sky distribution of galaxies across ALFAZOA Shallow, with the smaller dots representing sources from HyperLEDA and the larger those from our survey. The colors represent slices of velocity in Figure \ref{lsslb} and slices of longitude in Figure \ref{lsspolar}. The survey's ability to uncover the LSS across the Galactic plane is visible in these Figures.

ALFAZOA confirms the continuation across the ZOA of much of the structure predicted by \cite{Erdogdu06}, but also contradicts some of the predictions. We find that out of the 7 named large scale structures predicted to exist in this part of the ZOA (\citealt{Erdogdu06})-- three overdensities and four voids-- we can confirm the extent of two of the overdensities and one void, and find an extension of the third overdensity where it was not predicted. We will go into more detail on these in section \ref{sec:predicted}. Additionally, we find two unpredicted overdensities, see Section \ref{new} for more details.

\subsection{Predicted Structures} \label{sec:predicted}

\cite{Erdogdu06} created density reconstruction maps out to 16,000 \kms~from 2MRS data (\citealt{Huchra12}), using an expansion of the fields in Fourier-Bessel functions. Because of the impact of the ZOA on 2MRS (see lack of literature galaxies in the ZOA in Figures \ref{lsslb} and \ref{lsspolar}, and previous discussion in Section \ref{sec:counter}), \cite{Erdogdu06} had to extrapolate from the structure above and below the plane to fill in the regions from -10\degree $\leq b \leq$ 10\degree. Because of this, these density reconstruction maps are likely to make incorrect assumptions about the galaxy distribution in the ZOA, creating structure or misplacing expected structure. The location and extent of the predicted structures are defined in \cite{Erdogdu06}, and their borders are defined as the region where the density fluctuations are greater than 0 for overdensities, and $-$1 for voids.

\begin{figure*}
\begin{center}
\includegraphics[width=0.9\textwidth]{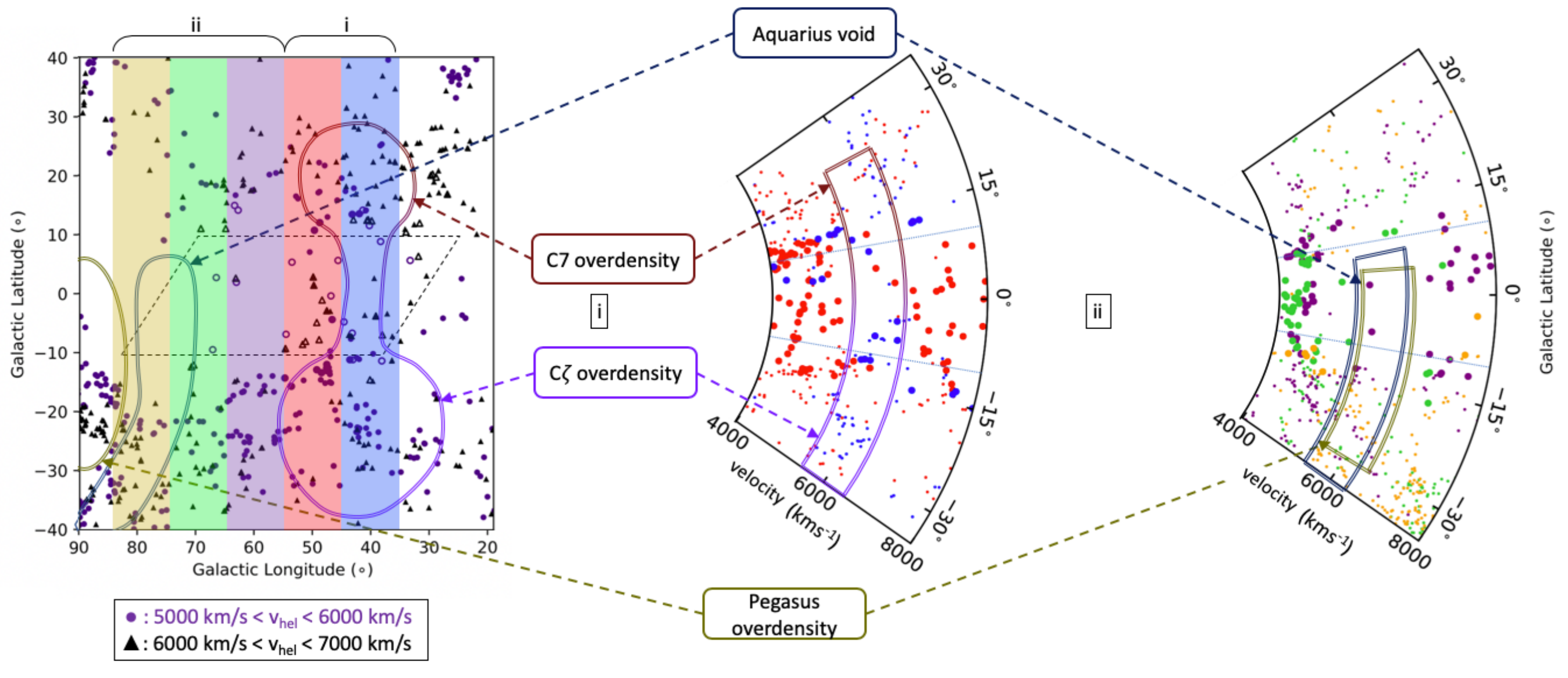}
\caption{Same as in Figure \ref{lss2000}, but with (left) velocities in the range 5000 \kms $\leq$ \vhel~$\leq$ 7000 \kms, and (middle and right) velocity ranging from 4000 \kms to 8000 \kms, with an additional 1000 \kms added at both ends. Of the four structures shown here, three are inside of the survey area (C7, C$\zeta$ and the Aquarius void). }
\label{lss6000}
\end{center}
\end{figure*}

\begin{figure*}[htbp]
\begin{center}
\includegraphics[width=0.9\textwidth]{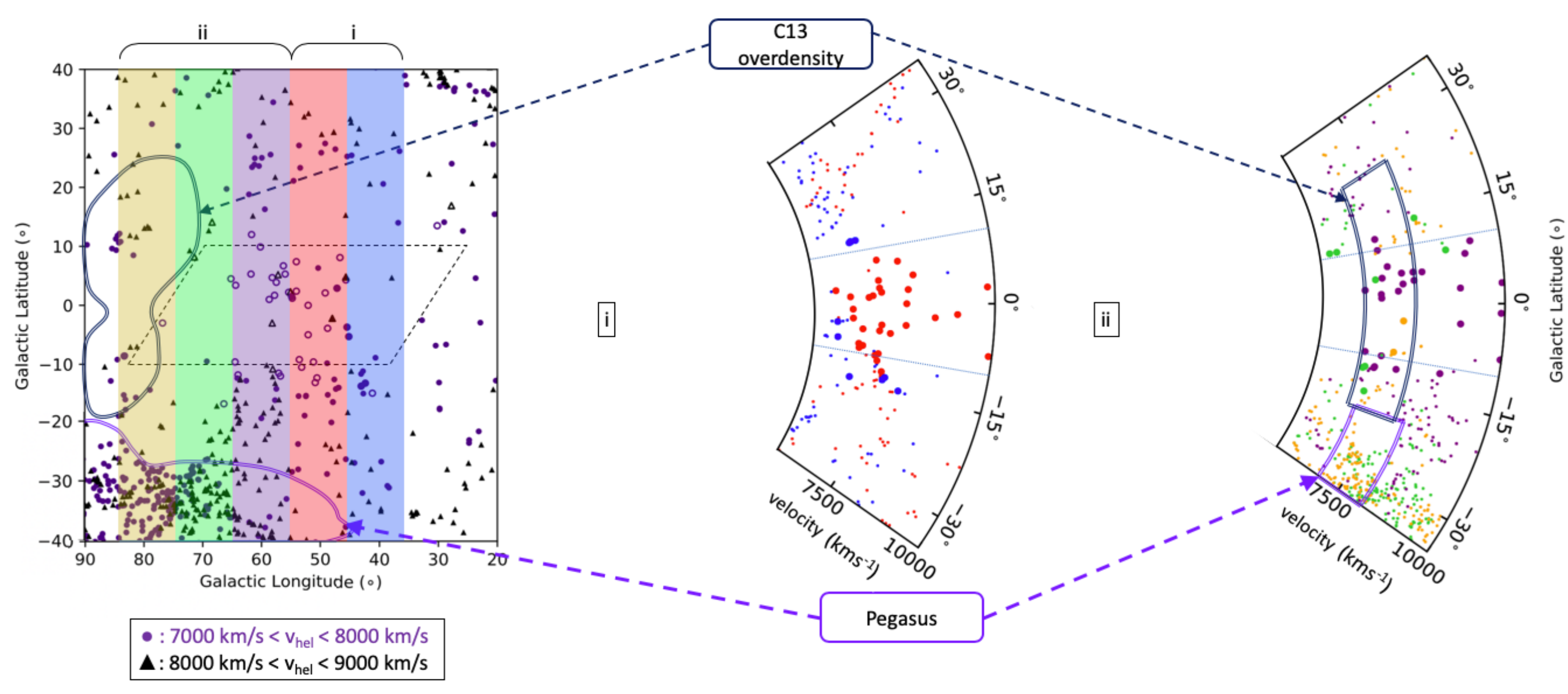}
\caption{Same as in Figure \ref{lss2000}, but with (left) velocities in the range 7000 \kms $\leq$ \vhel~$\leq$ 9000 \kms, and (middle and right) velocity ranging from 6000 \kms to 10,000 \kms, with an additional 1000 \kms added at both ends. Of the two structures shown here, only the C13 overlaps with edges of our survey area. }
\label{lss8000}

\bigskip

\includegraphics[width=0.9\textwidth]{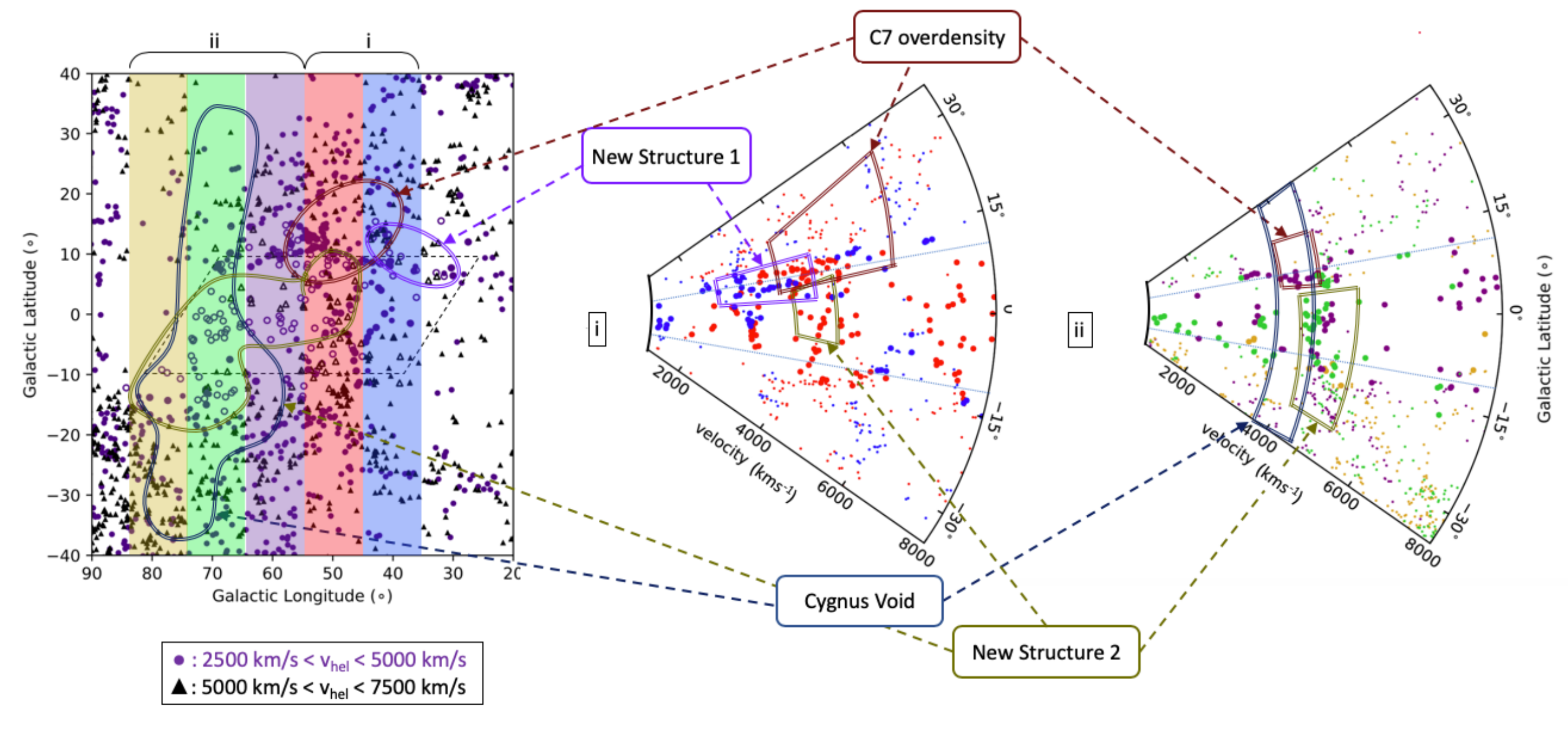}
\caption{Same as in Figure \ref{lss2000}, but with (left) velocities in the range 2500 \kms $\leq$ \vhel~$\leq$ 7500 \kms, and (middle and right) velocity ranging from 1000 \kms to 8000 \kms. Of the four structures shown here, two are the previously mentioned C7 and Cygnus Void and the other two are the newly identified structures. }
\label{lssnew}
\end{center}
\end{figure*}

Comparing ALFAZOA detections with the \cite{Erdogdu06} density maps allows us to check the effectiveness of predicting LSS in the ZOA. Figures \ref{lss2000}, \ref{lss4000}, \ref{lss6000} and \ref{lss8000} show plots of ALFAZOA detections alongside HyperLEDA galaxies with two different projections (one onto the galactic plane, and two others as wedge plots projected onto latitude) and indicate major overdensities and voids from the density reconstructions of \cite{Erdogdu06}, represented by colored outlines with double lines. 

\cite{Donley05} looked at galaxies in the ZOA at $l$=36\degree~$-$~52\degree. Our results are consistent with the overdensities they observed at 3000 \kms~and 4500 \kms.

\textbf{Delphinus Void:}
Figure \ref{lss2000} shows the distribution of ALFAZOA and LEDA galaxies in the region where the Delphinus Void should be found in the velocity range around 2000 \kms, as described in \cite{Erdogdu06}. The Delphinus void is the largest underdensity in the sky at 2500 \kms, and ALFAZOA Shallow only overlaps with the edges of it (around 3\% of the full predicted area). Only one ALFAZOA galaxy is found in this region, consistent with the predicted underdensity. Furthermore, the galaxies found at the edge of this region confirm the end of this void and that it does not extend into the ZOA.

\textbf{Corona Borealis Void:}
The predicted extent of the Corona Borealis void only  intersects with a corner of our survey area (see Figure \ref{lss4000}). We detected very few galaxies detected at its predicted position around 3500 \kms $\leq$ \vhel~$\leq$ 4500 \kms (only 6 blue galaxies in the outlined void region in the middle panel of Figure \ref{lss4000}). If we compare the middle panels of Figures \ref{lss2000} and \ref{lss4000}, we can note a new overdensity extending from about 1500 \kms~at $(l,b) \sim$ (30\degree,8\degree) to 4500\kms, at $(l,b) \sim$ (42\degree,10\degree), which joins the C7 overdensity at this end and extends across the Corona Borealis void. Therefore, we can only confirm the void at longitudes and latitudes lower than $(l,b) \sim$ (42\degree,10\degree).

\textbf{Cygnus Void:}
The largest departure from the predicted structure of the ZOA can be found with respect to the Cygnus Void. Figures \ref{lss4000} and \ref{lss6000} show a structure that bisects the predicted void close to the Galactic plane and reaches through to connect the C7 overdensity from around 4000 \kms~at $(l,b) \sim$ (55\degree,13\degree) up to the Pegasus overdensity at around 5500 \kms~and $(l,b) \sim$ (88\degree,-15\degree). There is no void at this predicted location and velocity range, as most clearly illustrated in the left panel of Figure \ref{lss4000}.

\textbf{C7 and C$\zeta$ Overdensities:}
We can confirm that the C7 overdensity is where predicted, from 3000 \kms~to 6000 \kms~at $(l,b) \sim$ (55\degree,10\degree), including an extension into the ZOA down to a latitude of $\sim$5\degree, by examining Figures \ref{lss4000} and \ref{lss6000} which shows the extension in longitude and velocity of this structure. The C$\zeta$ overdensity is also where predicted, although its extension goes farther into the ZOA at lower velocities, around 4500 \kms~and $(l,b) \sim$ (45\degree,-15\degree), where in Figure \ref{lss4000} the extension of the green segment in the middle can connect upwards through the red galaxies. As noted earlier, there is a new structure which shows an overdensity that spreads out from $(l,b) \sim$ (30\degree,8\degree) at 1500 \kms~and which connects to C7 at $(l,b) \sim$ (42\degree,10\degree). The predicted connection through the Galactic plane between C7 and C$\zeta$ is also confirmed, although it appears at a lower velocity than expected (around 4500 \kms~rather than 6000 \kms).

\textbf{Pegasus Overdensity:}
ALFAZOA only has a small overlap with Pegasus (seen in Figure \ref{lss6000}), which is confirmed as an overdensity from 4000 to 5000 \kms at $(l, b)$ = (80\degree, -15\degree ). It lies beyond $l$=90\degree~in the velocity range from 3000 \kms~to 5000 \kms. We do have a structure that connects the C7 overdensity from around 4000 \kms~and $(l,b) \sim$ (55\degree,13\degree) up to the Pegasus overdensity at around 5500 \kms~and $(l,b) \sim$ (88\degree,-15\degree), discussed earlier. The Pegasus overdensity continues at higher velocities to spread further away from the ZOA as predicted, and is shown in Figure \ref{lss8000}.

\textbf{Aquarius Void:}
The Aquarius void is confirmed at $l$=75\degree, from $b$=-15\degree to $b$=5\degree, and \vhel=5000 to 7000 \kms. A predicted small unnamed void at $l$ = 60 \kms~is confirmed over the same region. This can be seen in Figure \ref{lss6000}.

\textbf{C13 Overdensity:}
ALFAZOA only has a small overlap with the edges of the C13 overdensity (see Figure \ref{lss8000} ), in a region which corresponds to the A field in the survey. Due to problems described in section \ref{sec:ad}, there were very few detections at this velocity range, so we can't say much about this region.

\subsection{New Structures} \label{new}

We identify two new structures in the region covered by the survey, illustrated in Figure \ref{lssnew}. The first is the small one which extends from $(l,b) \sim$ (30\degree,8\degree) at 2500 \kms~and connects to C7 at $(l,b) \sim$ (42\degree,10\degree) at 4500 \kms. The second one also connects to C7, but links up with the Pegasus overdensity across the non-existent Cygnus void in the ZOA, at around 4500 \kms. This would then connect with the larger Perseus-Pisces extension described by \cite{Kraan18}, which is a large chain that was predicted to pass through the Galactic Plane at around $l \sim$ 80\degree~(where the second new structure connects to the Pegasus overdensity), and joins up with the C7 overdensity at $(l, b) \sim$ (50\degree, +15\degree) at around 5000 \kms.


\section{Conclusions} \label{sec:concl}

We presented the data from the blind \HI~survey ALFAZOA Shallow carried out with the Arecibo 305m telescope with the 7 beam ALFA receiver. This survey covered a section of the ZOA in the northern sky, at 30\degree $\leq l \leq$ 75\degree~and $ \lvert b \lvert $ $\leq$ 10\degree~with full sensitivity, and an additional region of partial coverage at $ \lvert b \lvert $ $\leq$ 15\degree. The summary of the results reported here is as follows:

\begin{itemize}
 \item Out of the 403 detected sources, 64\% (259 galaxies) did not have existing redshift information. The survey provides new redshifts for these galaxies;
 \item In total, 58\% of galaxies in this survey had a possible counterpart in the optical, near-IR, or \HI~reported in the literature;
 \item The positional accuracy of unresolved \HI~sources is between  0.2' and 1.7', with a general value of 0.4', and the survey reached its expected sensitivity of rms = 5.4 mJy at 9 \kms~channel resolution in 45\% of the area (B and C fields), whereas it decreases to rms = 7 mJy at 20.3 \kms~channel resolution for the rest of the area (A and D fields);
 \item The survey confirms the extent of the predicted C7, C$\zeta$, and Pegasus overdensities, as well as the Aquarius void through the ZOA; 
 \item The Corona Borealis and Delphinus voids were found, but not where predicted, whereas the Cygnus void was not found at all;
 \item We also discovered two new, previously unsuspected structures, one starting at ($l, b, v $= 30\degree, +13\degree, 1500 \kms) and connecting with the C7 overdensity at ($l, b, v$ = 42\degree, +13\degree, 4500 \kms), the other one at ($l, b, v$ = 8\degree, -10\degree, 5000 \kms) at the Pegasus overdensity and connecting with the C7 overdensity at ($l, b, v$ = 42\degree, +13\degree, 4500\kms).
\end{itemize}

\section{Acknowledgements} \label{sec:ack}
The Arecibo Observatory is operated by SRI International under a cooperative agreement with the National Science Foundation (AST-1100968), and in alliance with Ana G. Mendez-Universidad Metropolitana, and the Universities Space Research Association.
Support for this work was provided by the NSF through the Grote Reber Fellowship Program administered by Associated Universities, Inc./National Radio Astronomy Observatory. The National Radio Astronomy Observatory is a facility of the National Science Foundation operated under cooperative agreement by Associated Universities, Inc. 
This research has made use of the NASA/IPAC Extragalactic Database (NED), which is operated by the Jet Propulsion Laboratory, California Institute of Technology, under contract with the National Aeronautics and Space Administration; the HyperLEDA database (http://leda.univ-lyon1.fr), and the NASA/IPAC Infrared Science Archive, which is operated by the Jet Propulsion Laboratory, California Institute of Technology, under contract with the National Aeronautics and Space Administration.

\bibliographystyle{hapj}
\bibliography{biblio.bib}

\end{document}